\documentclass{aa}  

\usepackage{txfonts}
\usepackage{graphicx}	
\usepackage{amsmath}	
\usepackage{amssymb}	
\usepackage{subcaption}
\usepackage{float}
\usepackage{enumerate}
\usepackage{multirow}
\usepackage[export]{adjustbox}
\usepackage{tabularx}
\begin{document} 

   \title{A new perspective on the stellar Mass-Metallicity Relation of quiescent galaxies from the LEGA-C survey.}
   \titlerunning{A new perspective on the stellar MZR}

   \author{Davide Bevacqua\inst{1,2}\thanks{E-mail: davide.bevacqua@inaf.it}
          \and
          Paolo Saracco\inst{1}
          \and
          Alina Boecker\inst{3,4}
          \and
          Giuseppe D'Ago\inst{5}
          \and
          Gabriella De Lucia\inst{6}
          \and
          Roberto De Propris\inst{7,8}
          \and
          Francesco La Barbera\inst{9}
          \and
          Anna Pasquali\inst{10}
          \and
          Chiara Spiniello\inst{11,9}
          \and
          Crescenzo Tortora\inst{9}}

   \institute{INAF - Osservatorio Astronomico di Brera, via Brera 28, 20121 Milano, Italy              
         \and
             DiSAT, Universit\'{a} degli Studi dell'Insubria, via Valleggio 11, I-22100 Como, Italy
          \and
          Instituto de Astrof\'{i}sica de Canarias (IAC), c/ V\'{i}a L\'{a}ctea s/n, La Laguna 38205, Spain
          \and
          Departamento de Astrof\'{i}sica, Universidad de La Laguna (ULL), Av. Astrof\'{i}sico Francisco S\'{a}nchez s/n 38200 La Laguna, Spain 
          \and
           Institute of Astronomy, University of Cambridge, Madingley Road, Cambridge CB3 0HA, United Kingdomà
           \and
           INAF - Osservatorio Astronomico di Trieste, via G.B. Tiepolo 11, I-34143 Trieste, Italy
           \and
           FINCA, University of Turku, Vesilinnantie 5, Turku, 20014, Finland
           \and
           Department of Physics and Astronomy, Botswana International University of Science and Technology, Private Bag 16, Palapye, Botswana
           \and
           INAF -  Osservatorio Astronomico di Capodimonte, Via Moiariello  16, 80131, Naples, Italy
           \and
			Astronomisches Rechen-Institut, Zentrum f\"{u}r Astronomie der Universit{\"a}t Heidelberg, M\"{o}nchhofstrasse 12 - 14, 69120 Heidelberg, Germany
			\and
			Sub-Dep. of Astrophysics, Dep. of Physics, University of Oxford, Denys Wilkinson Building, Keble Road, Oxford OX1 3RH, UK
           }

   \date{Received 04 December 2023 / Accepted 15 July 2024}

 
  \abstract
	{We investigate the stellar Mass-Metallicity Relation (MZR) using a sample of 637 quiescent galaxies with $10.4 \leq \log($M$_*$/M$_\odot) < 11.7$ selected from the LEGA-C survey at $0.6 \leq z \leq 1$. We derive mass-weighted stellar metallicities using full-spectral fitting. We find that while lower-mass galaxies are both metal -rich and -poor, there are no metal-poor galaxies at high masses, and that metallicity is bounded at low values by a mass-dependent lower limit. This lower limit increases with mass, empirically defining a MEtallicity-Mass Exclusion (MEME) zone. We find that the spectral index MgFe $\equiv \sqrt{\mbox{Mgb}\times\mbox{Fe4383}}$, a proxy for the stellar metallicity, also shows a mass-dependent lower limit resembling the MEME relation. Crucially, MgFe is independent of stellar population models and fitting methods. By constructing the Metallicity Enrichment Histories, we find that, after the first Gyr, the Star Formation History of galaxies has a mild impact on the observed metallicity distribution. Finally, from the average formation times, we find that galaxies populate differently the metallicity-mass plane at different cosmic times, and that the MEME limit is recovered by galaxies that formed at $z\geq3$. Our work suggests that the stellar metallicity of quiescent galaxies is bounded by a lower limit which increases with the stellar mass. On the other hand, low-mass galaxies can have metallicities as high as galaxies $\sim 1$ dex more massive. This suggests that, at $\log($M$_*$/M$_\odot)\geq 10.4$, rather than lower-mass galaxies being systematically less metallic, the observed MZR might be a consequence of the lack of massive, metal-poor galaxies.}   


   \maketitle
%

\section{Introduction}\label{sect:intro}
\nolinenumbers

The elemental abundance of galaxies holds fundamental information on how they formed and evolved. In particular, stellar metallicity is primarily driven by stellar nucleosynthesis, and the final metal content of a galaxy is tightly related to its Star Formation History (SFH). Indeed, the metallicity of a galaxy is the result of the interplay between how quickly its stars produce and release metals in the interstellar medium (ISM), and how much of these metals are retained and reprocessed into new stars.

In the local Universe, galaxies exhibit a positive correlation between their stellar mass and metal content of both stars and gas \citep{Tremonti+04, Thomas+2005, Gallazzi+05,  Panter+08, Thomas+10, Gallazzi+14, Choi+14, McDermid+15, Ginolfi+20}. This is known in the literature as the Mass-Metallicity Relation (MZR)\footnote{In this paper, we refer to the stellar MZR, unless otherwise specified.}, which indicates that more massive galaxies are, on average, more metal-rich than less massive ones. 

Being the stellar mass a proxy of the gravitational potential, the common interpretation of the MZR is that, because of their deeper potential well, more massive galaxies are capable of retaining more metals against the galactic outflows, and reprocessing them into new stars, in contrast to lower mass galaxies that have shallower potential wells \citep{Tremonti+04, Chisholm+18, D'Eugenio+18, Barone+20, ppxf_2023}. However, the origin of the MZR is still debated, because of the many mechanisms involved in the interplay between outflows, inflows, and enrichment rate  \citep{Finlator_Dave_2008, Spitoni+10, Spitoni+17, Dave+11}, as well as star formation efficiency \citep{Calura+09}, and the Initial Mass Function (IMF) \citep{Conroy+12, Spiniello+12, spider, MN+15}, determining the final metal content of a galaxy.

Although both quiescent and star-forming galaxies show a positive correlation of metallicity with mass, it has been shown that they follow two different MZRs. This has been interpreted as the consequence of the suppression of the star formation, in quiescent galaxies, due to a halt of the external gas supply, namely the so-called `starvation' \citep{Peng+15, Trussler+19}. However, the differences in the metal content of quiescent and star-forming galaxies could also be explained in terms of the interplay between different gas infall time-scales and outflows \citep{Spitoni+17}. Further, the two different MZRs are linked to structural differences, and the local relation between metallicity and surface mass density \citep{ZG22}.

A major issue related to the study of the metallicity of local galaxies is the difficulty of reconstructing the full star formation and assembly history of galaxies. Indeed, events like mergers, and gas accretion can significantly affect the metallicity of galaxies. Further, galaxies formed at different cosmic times can have different stellar population properties, thus affecting the overall distribution of metallicities. All these effects combined make the interpretation of the SFHs non-trivial, and the relation between mass and metallicity less obvious. Observationally, the easiest way to mitigate this issue is to study galaxies at higher redshifts, when the time available for cosmic events to take place is shorter, and, consequently, the impact on the overall distribution of the stellar population properties is milder. 

At intermediate redshifts, studies on the stellar metallicity of quiescent galaxies have revealed different results. For example, \cite{Gallazzi+14} find a MZR for quiescent galaxies at $z \sim 0.7$ consistent with the one for local ETGs \citep{Gallazzi+05}, as also confirmed, e.g., by \cite{Choi+14}, and \cite{Saracco+23} at even higher redshift ($1\leq z\leq 1.4$). On the other hand, e.g., \cite{Leethochawalit_2019}, and \cite{Carnall+19, Carnall+22} find that, at fixed mass, the average metallicity of quiescent galaxies is lower at higher redshift than their local counterpart, indicating an evolution of the MZR. Other studies on smaller samples of quiescent galaxies, using single or stacked spectra, also provide a variety of results at comparable masses \citep{Lonoce+14, Onodera+15, Kriek+16, Saracco+19, Kriek+19, Saracco+20, Lonoce+20, Carnall+22}. However, the different results obtained by different studies could stem from differences in the methods to estimate metallicity\footnote{For example, using similar samples from the same galaxy survey, \cite{Saracco+23} and \cite{Carnall+22} find significantly different results, using different methods.}, as well as the definition itself of metallicity (total metallicity, [Z/H], or iron content, [Fe/H]).

An evolution of the stellar MZR might be the consequence of cosmic evolution affecting more heavily the stellar population properties of lower mass galaxies, than higher mass ones \citep{Maiolino+08, Fontanot+09}. This would imply a larger range of metallicities for lower mass galaxies, as it is actually observed. The observed scatter in metallicity might also reflect the rich variety of SFHs and mass assembly histories of galaxies \citep{Tacchella+22}. In this sense, studying only the average trend, i.e. the MZR, could be preventing us from obtaining a clear picture of how galaxies of a given mass are enriched in metals, since average values are not fully representative of the complex mechanisms determining and affecting the final metallicity of galaxies. Therefore, studying in more detail the overall metallicity distribution can provide valuable information to better characterize the relation between metallicity and mass.

In this paper, we study the relation between stellar metallicity and stellar mass for a large sample of quiescent galaxies selected from the Large Early Galaxy Astrophysics Census survey (\citealt{legac}, hereafter, LEGA-C), and its dependence on the SFH. The stellar metallicity and SFH of LEGA-C galaxies has already been investigated by several works \citep{Beverage+21, Barone+22,Borghi+22,  Beverage+23, Wu+18, Chauke+18, Chauke+19, Sobral+22, ppxf_2023}.

This paper is structured as follows. In section \ref{sect:data} we describe the data and selection criteria used to build the sample. In section \ref{sect:methods} we describe the methods used to estimate the stellar population parameters, as well as the results of the fits. In section \ref{sect:mzr} we describe the relation between metallicity and mass found for our sample of quiescent galaxies, and discuss a new view of the MZR. In section \ref{sect:cosmicevol} we evaluate the impact of cosmic evolution on metallicity, both in terms of SFH, and different formation epochs. Finally, in section \ref{sect:summary} we summarize our results and discuss them in section \ref{sect:discussion}.

Throughout this paper we adopt a flat $\Lambda$CDM cosmology with H$_0 = 70$ km s$^{-1}$ Mpc$^{-1}$ and $\Omega_{\rm m} = 0.3$.

\section{Data and sample selection}\label{sect:data} 

\begin{figure}
\includegraphics[width=\columnwidth]{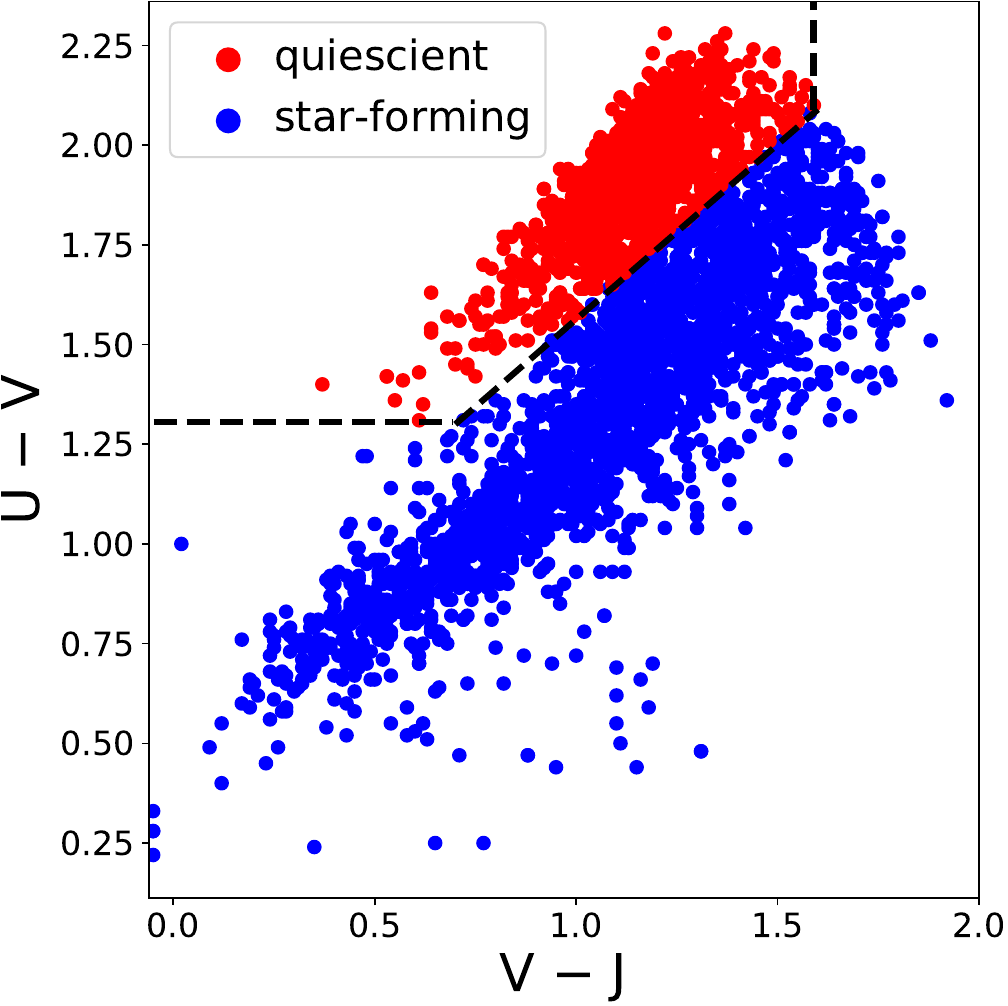}
\caption{UVJ diagram of LEGA-C galaxies in the DR3. The black dashed line marks the empirical separation between quiescent and star-forming galaxies (equation 4 of \citealt{UVJ}).}
\label{fig:uvj}
\end{figure}

We study a large sample of quiescent galaxies selected from LEGA-C. LEGA-C is a spectroscopic survey of galaxies observed at intermediate redshifts ($z \sim 0.6 - 1.0$) in the Cosmological Evolution Survey (COSMOS) field \citep{cosmos}, using the VIsible Multi-Object Spectrograph \citep{vimos} on the Very Large Telescope (VLT). The survey observed 4209 galaxies, selected from the UltraVISTA catalog \citep{ultravista}, reaching an approximate signal-to-noise (S/N) of about 20 \AA$^{-1}$ in the continuum. We use integrated spectra from the third Data Release (\citealt{DR3}; hereafter, DR3), with a nominal spectral resolution of R $\approx$ 2500 \citep{Straatman+18} and observed wavelength range $6300 \mbox{\AA} - 8800 \mbox{\AA}$.

In this work, we focus only on quiescent galaxies, namely those galaxies that show no evidence of ongoing star formation, and hence have stopped forming stars. 

We first select them using the photometric UVJ diagnostic diagram (e.g., \citealt{UVJ}), which is shown in Fig. \ref{fig:uvj} for all LEGA-C galaxies. Red points in the plot are classified as quiescent galaxies and are kept from the input catalog. Then, we further clean the sample using the LEGA-C flags and exclude galaxies with irregular morphologies (FLAG\_MORPH = 1), galaxies with light coming from different redshifts (FLAG\_MORPH = 2), and galaxies with bad flux calibration (FLAG\_SPEC = 2) (see \citealt{DR3} for details). Finally, we exclude galaxies with S\'{e}rsic index n$_{\rm sers}$ < 2.5 and galaxies with axial ratio q$_{\rm ax}$ < 0.3. These two last criteria aim at minimizing (maximizing) the number of late- (early-)type galaxies in our sample.

To study a possible evolution in redshift, we divide the sample into four redshift bins, from $z = 0.6$ to $z = 1.0$ at step of $\Delta z = 0.1$, and exclude all galaxies outside this range, thus covering about 2 Gyr of evolution (the Universe was about 5 Gyr old at $z = 1$ and 7 Gyr old at $z = 0.6$) at step of about 0.5 Gyr. To take into account the mass completeness limits of LEGA-C, we adopt a conservative choice and exclude from our sample all galaxies with mass lower than the completeness limit at the highest redshift of each redshift bin considered. More specifically, we exclude galaxies with masses lower than $\log_{10}$(M$_*$/M$_\odot) = 10.4,\, 10.6, \, 10.8, \, \mbox{and}\, 11$ in the first ($ 0.6 \leq z < 0.7$), second ($ 0.7 \leq z < 0.8$), third ($ 0.8 \leq z < 0.9$) and last ($ 0.9 \leq z \leq 1$) redshift bin, respectively.

The final sample consists of 637 quiescent galaxies. The redshifts and spectral indices are provided by the DR3. We then estimate the stellar masses by fitting the UltraVISTA photometry \citep{ultravista} with the spectroscopic redshifts provided by LEGA-C, using the C++ implementation of the \texttt{FAST} code\footnote{Available from \url{https://github.com/cschreib/fastpp}} \citep{fast}. To perform the fit, we use models by \cite{BC03} with solar metallicity, assuming a delayed exponentially declining SFH, a \cite{Chabrier_IMF} IMF, a \cite{KC13} dust attenuation law.

\section{Age and metallicity estimates from full-spectral fitting}\label{sect:methods}

\begin{figure}
\includegraphics[width=\columnwidth]{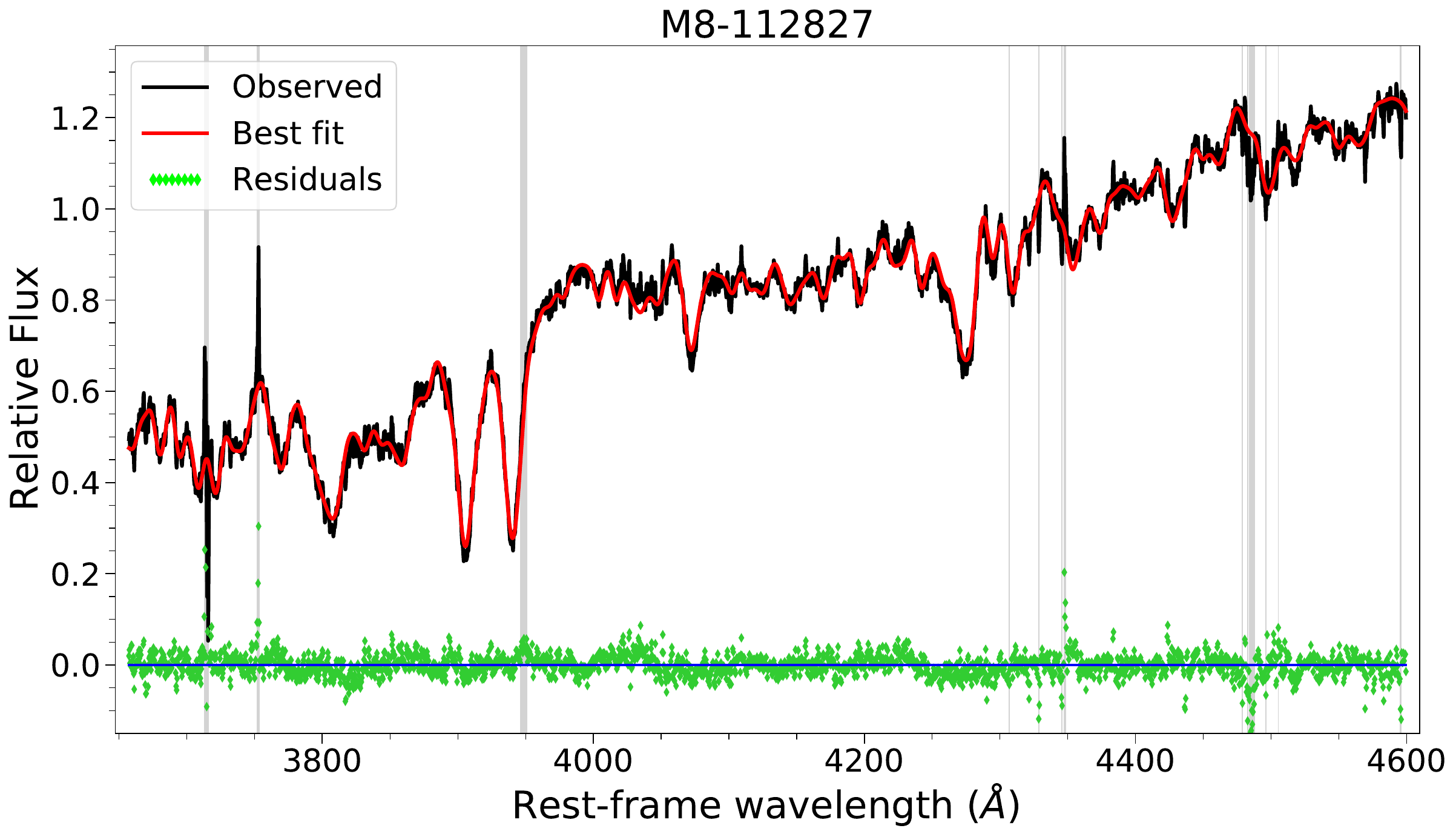}
\caption{Example of a fit performed with \texttt{pPXF} on the galaxy with LEGA-C ID M8-112827. The black line is the observed spectrum. The red line is the best-fit model. The green diamonds are the residuals, whose median value is indicated by the blue horizontal line. The grey shaded lines are the masked regions.}
\label{fig:fitexample}
\end{figure}

\begin{table}
	\centering
	\caption{Estimated stellar population parameters}
	\label{tab:stellarpop}
	\begin{tabularx}{\columnwidth}{>{\centering\arraybackslash}p{1.8cm}>{\centering\arraybackslash}p{1.3cm}>{\centering\arraybackslash}p{1cm}>{\centering\arraybackslash}p{1.3cm}>{\centering\arraybackslash}p{1.3cm}}
		\hline
		ID LEGA-C & Age & [M/H] & $t_{\rm form}$ & $\Delta\tau$\\
		 & (Gyr) & (dex) & (Gyr) & (Gyr)\\
		(1) & (2) & (3) & (4) & (5)\\
		\hline
		& & & &\\
		M16-38110 & 6.50 & $-0.10$ & 0.22 & 0.40 \\
		M17-40011 & 3.79 & $-0.06$ & 3.46 & 3.70 \\
		M14-41209 & 4.80 & $-0.12$ & 2.09 & 0.40 \\
		... & ... & ... & ... & ...\\
		& & & & \\
		\hline
	\end{tabularx}
	\begin{minipage}{\columnwidth}
	The complete table is available at the CDS. Columns: (1) ID LEGA-C of the file associated with the spectrum. (2) Mass-weighted age estimated from fits, using equation \eqref{eq:age}; we assume a fixed error of 0.07 dex for all galaxies (see appendix \ref{app:errors}). (3) Mass-weighted metallicity estimated from fits, using equation \eqref{eq:met}; we assume a fixed error of 0.06 dex for all galaxies (see appendix \ref{app:errors}). (4) Average formation time estimated using equation \eqref{eq:tform}. (5) $\Delta\tau \equiv \tau_{95} - \tau_{5}$, where $\tau_x$ is the time at which a galaxy formed $x$-percent of its mass, as measured from its SFH (see section \ref{sect:SFH}).
	\end{minipage}

\end{table}

We estimate the mass-weighted ages and metallicities by fitting the LEGA-C spectra using the penalized pixel fitting (\texttt{pPXF}) method and code described in \citet[updated to version v8.2.2]{ppxf_2004, ppxf_2017, ppxf_2023}. The code performs the full-spectral fit by linearly combining template spectra of different ages and metallicities. The resulting composite best-fit model is the one minimizing the $\chi^2$.

As templates, we use the E-MILES Simple Stellar Population (SSP) models \citep{EMILES}, which are entirely based on observed stars. More specifically we use models with BaSTI isochrones \citep{basti} and a Chabrier IMF \citep{Chabrier_IMF}. We restrict to the safe ranges described in \cite{EMILES}, namely we use only models with metallicities [M/H] $= -1.79, -1.49, -1.26, -0.96, -0.66, -0.35,$ $-0.25, +0.06, +0.15, +0.26$ and ages $\geq 0.11$ Gyr. 

As the upper limit of the SSPs' age, we consider a maximum input age of  6.5, 6, 5.5, and 5 Gyr for galaxies at redshift $0.6 \leq z < 0.7$, $0.7 \leq z < 0.8$, $0.8 \leq z < 0.9$, and $0.6 \leq z < 1.0$ respectively. In Appendix \ref{app:age_ssp} we show that including SSPs up to 1.5 Gyr older has a negligible effect on the metallicity estimates. In this case, the output ages become slightly older, but in most cases consistent within the estimated errors (Appendix \ref{app:errors}). However, since some cases may depend on the choice of the input SSP ages, throughout this paper we discuss how this choice affects our results, when needed. Finally, including much older SSPs, instead, often provides unphysical outputs, namely age estimates much older than the age of the Universe. This is likely due to the degeneracy between age and metallicity.

We perform the fit of the spectra as follows. Hence, we perform two fits: from the first fit, we get the residuals between the galaxy and the best-fitting model, and make a robust estimate of the standard deviation of these residuals, $\sigma_{\rm std}$, that we use to mask all spectral pixels deviating more than $3\sigma_{\rm std}$. Then, the second fit provides the best-fitting model. The kinematic broadening is taken into account during the fit by \texttt{pPXF}.

Each fit is performed using both multiplicative polynomials of degree 4 and a Calzetti reddening curve \citep{Calzetti_2000}, over the rest-frame spectral range 3600 - 4600 \AA . In Appendix \ref{app:fittest} we discuss these choices in detail. Briefly, by simulating LEGA-C galaxies from E-MILES models, we find that the combination of a reddening curve and low-order polynomials generally provides estimates of age and metallicity closest to the input values. Then, we choose the spectral range to be common to most galaxies, given the large range of redshifts considered here. We verify that shortening the range by 200 \AA \; to the redder or bluer part of the spectrum has a negligible effect on our estimates. Since we find that extending the fit up to 5200 \AA \; (reached by < 1/3 of galaxies) can affect the age estimates, while not changing metallicity estimates, when necessary, we discuss the effect of extending the fitting range on our results.

We also fit the gas emission lines (modeled as gaussians). In particular, we fit the Balmer series, for which we fix the flux ratio (\texttt{tie\_balmer = True}), and the [OII] doublet. In Fig. \ref{fig:fitexample} we show an example of the fit performed on a LEGA-C galaxy.

The average mass-weighted ages and metallicities are calculated as:

\begin{equation}\label{eq:age}
{\rm log}_{10}{\rm Age} = \frac{\sum_i  w_i \rm{log_{10} Age}_i}{\sum_i w_i}
\end{equation}

\begin{equation}\label{eq:met}
{\rm [M/H]} = \frac{\sum_i w_i {\rm [M/H]}_i}{\sum_i  w_i}
\end{equation}

\noindent where $w_i$ is the weight of the $i$-th template and the sums are performed over all the input templates.

Since the redshift range covered by LEGA-C galaxies corresponds to an age difference of about 2 Gyr, to make results comparable for the whole sample, we rescale the ages to the age of the Universe at the observed redshifts. We thus define the cosmic formation time, $t_{\rm form}$, relative to the mass-weighted age of a galaxy as:
\begin{equation}\label{eq:tform}
        t_{\rm form} \equiv \mbox{Age}_{\rm U} (z) - \mbox{Age \; ,}  
\end{equation}
where Age$_{\rm U} (z)$ is the age of the Universe at the redshift of the galaxy, and Age is the age of the galaxy estimated with \texttt{pPXF}.

Using $t_{\rm form}$ instead of ages allows for direct comparisons among galaxies, at all redshifts, since we are virtually re-arranging the timeline of galaxies to have the same zero-point, namely the age of the Universe.

Our estimates of stellar parameters are reported in Table \ref{tab:stellarpop}, and provided as supplementary material to this paper.

\section{The Metallicity-Mass diagram of LEGA-C quiescent galaxies}\label{sect:mzr}

\subsection{The Metallicity-Mass diagram}

\begin{figure}
\includegraphics[width=\columnwidth]{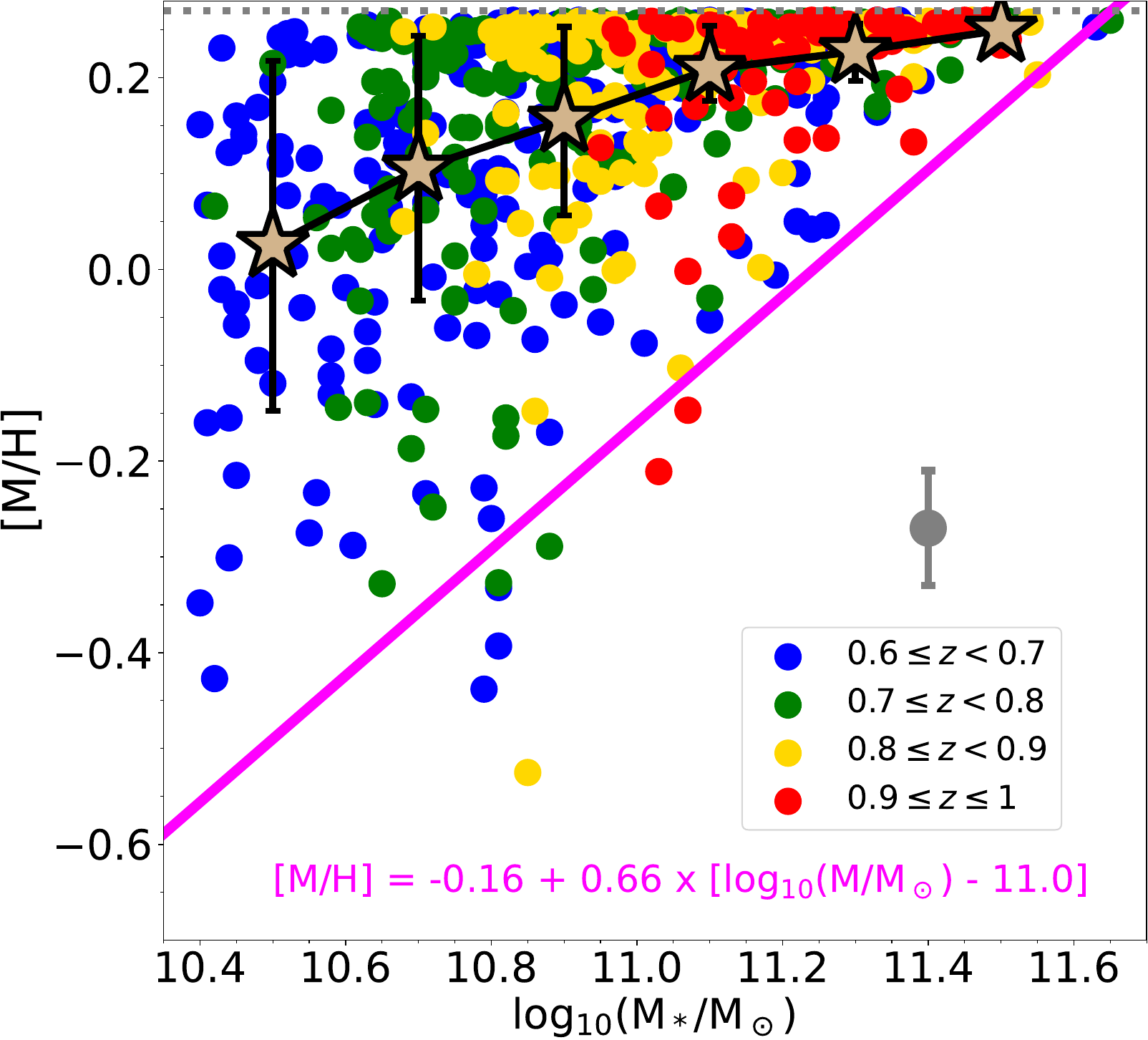}
\caption{Metallicity-Mass diagram of our sample of quiescent galaxies. Colors indicate different redshift bins, as described in the legend. The grey point indicates the typical error on [M/H]. The stars are the average metallicity values at different mass bins, regardless of the redshift bins, and the associated error bars are the corresponding 16th and 84th percentiles. The solid magenta line is the MEME relation, corresponding to the linear fit containing 99$\%$ of the galaxies in each mass bin.}
\label{fig:meme}
\end{figure}

In Fig. \ref{fig:meme} we show the distribution of galaxies on the metallicity-mass diagram and color-code them according to the redshift bins. As the figure shows, the distribution of metallicities does not change among different redshift bins\footnote{By performing KS-tests on all the combinations of the four samples, taking into account the mass completeness limit of the different redshift bins, we always get p-values $>0.5$}. For this reason, from now on we study the metallicity of our sample assuming it is not affected by redshift. 

In Fig. \ref{fig:meme} we highlight how the average metallicity increases with the stellar mass (see star symbols), in agreement with the MZR observed for local galaxies. Further, the metallicity scatter decreases with increasing stellar mass.

However, we point out that the MZR is not fully representative of the overall distribution of metallicity we observe in Fig. \ref{fig:meme}. Indeed, while at high masses galaxies are all metal-rich, at lower masses they are both metal -rich and -poor\footnote{We verified that there is no correlation between the S/N and M$_*$ for the LEGA-C galaxies. Note that this behavior is also seen in simulations (Appendix \ref{app:simulations}), which are not affected by the S/N.}. Then, low mass does not necessarily imply low metallicity, as a simplistic interpretation of the MZR may suggest. 

Furthermore, we note that, as the mass increases, there is a lack of galaxies with low metallicities, empirically defining a zone of avoidance. As highlighted by the magenta line (defined in section \ref{sect:meme}), this lower limit in metallicity increases with mass. On the other hand, lower-mass galaxies can reach metallicity as high as the most massive galaxies (although fits are limited by the maximum metallicity of input models; below, we show that spectral indices, which are independent of SSP models, show the same result). Hence, we argue that the average increase of metallicity with stellar mass (i.e. the MZR) is a consequence of the mass-dependent lower limit in metallicity. In other words, in the mass range considered, the MZR arises not because the metallicity is typically low for low-mass galaxies and high for the massive ones, but because there are no high mass galaxies with low metallicity (see also \citealt{Saracco+23}).

\begin{figure}
\includegraphics[width=\columnwidth]{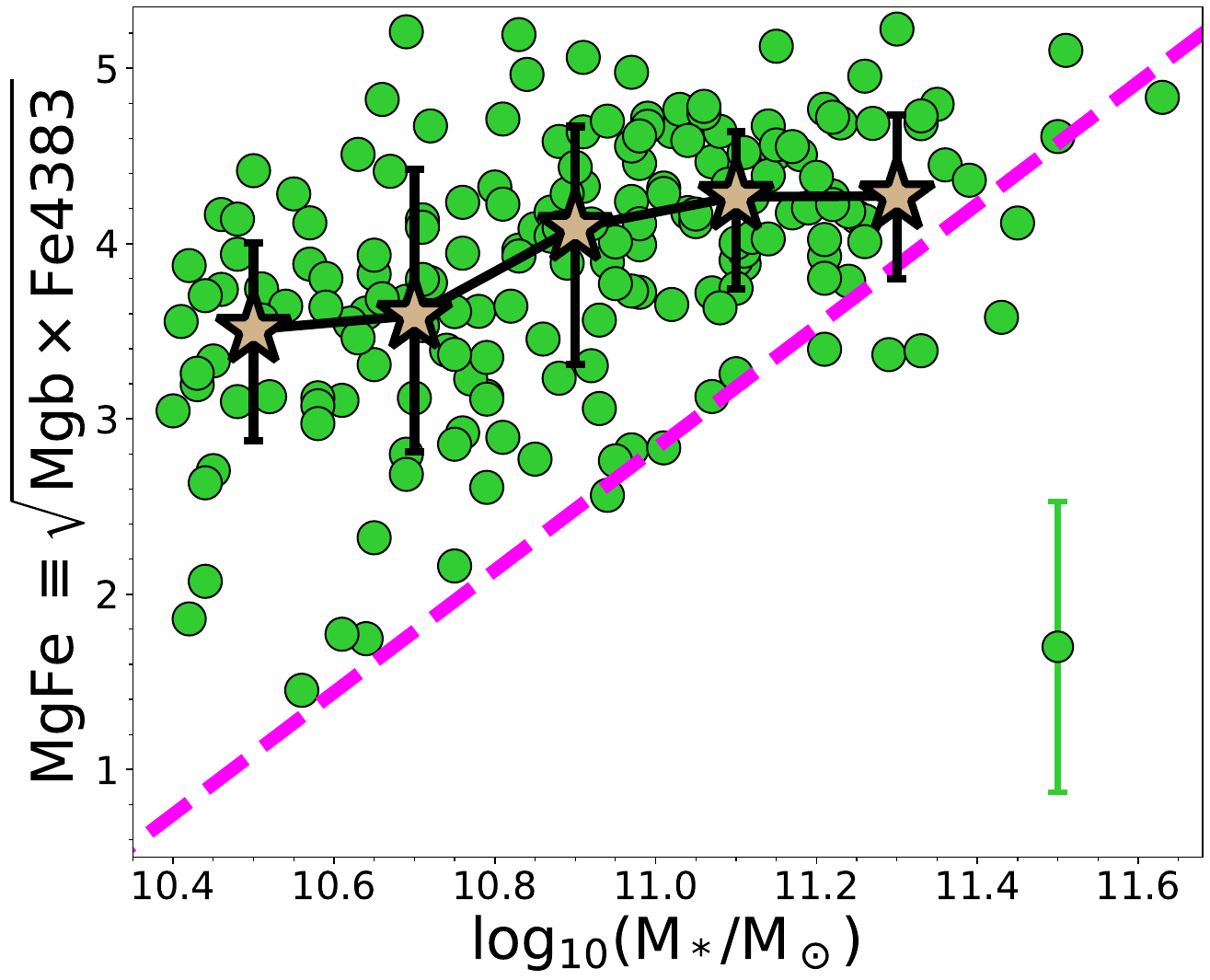}
\caption{MgFe-Mass diagram, where MgFe $\equiv\sqrt{{\rm Mgb} \times {\rm Fe4383}}$ is used as a proxy for the global metallicity. The point in the lower right corner represents the typical error on MgFe. The stars are the average metallicity values at different mass bins, and the associated error bars are the corresponding 16th and 84th percentiles. The dashed magenta line is the MEME relation estimated from indices: MgFe $= 2.83+3.48 \times [$log$_{10}$(M$_*$/M$_\odot) - 11]$.}\label{fig:meme_mgfe}
\end{figure}

\subsection{The MEME relation}\label{sect:meme}

To highlight the feature discussed above, we define a linear relation below which virtually no galaxies are found, as follows. First, we divide the sample into mass bins 0.2 dex wide, and sampled at 0.1 dex (thus bins are overlapping). Then, we apply the gaussian kernel density estimation of the metallicities at each mass bin and take the lower bound of the $99 \%$ confidence interval. Hence, we linearly interpolate the lower bounds of all mass bins, and perform a linear fit of the interpolated values using the least squares method that provides the slope and intercept of the following linear relation:
\begin{equation}\label{eq:meme}
    \mbox{[M/H]} = -0.16 + 0.66 \times [\log_{10}(\mbox{M}_*/\mbox{M}_\odot) - 11.0] \; .
\end{equation}
We call equation \eqref{eq:meme} the MEtallicity-Mass Exclusion (MEME) relation, and the region of the diagram below it the MEME zone. We plot the MEME relation in Fig. \ref{fig:meme} with a magenta solid line.

Another way of interpreting the MZR is by noticing that the metallicity range spanned by galaxies is narrower at higher masses, i.e. the observed metallicity scatter reduces as the mass increases, towards high metallicity values. We verified that the same is observed when considering only galaxies with high S/N. However, the [M/H] values are limited by models, and the crowding of galaxies at the highest fitted metallicity, [M/H] = 0.26 dex, observed in Fig. \ref{fig:meme} suggests that, if we had models with higher metallicities, galaxies would likely reach higher [M/H] values\footnote{We further checked that including models extrapolated up to [M/H] $ = 0.8$ dex does not change this result, with lower-mass galaxies reaching metallicities as high as the most massive galaxies; however, in this case, no galaxy reaches [M/H] = 0.8 dex.}. Therefore, to probe the observed metallicity distribution we need to rely on some other indicator, independent of models and fitting code.


To this end, we consider spectral indices. In particular, we use a combination of magnesium and iron as a proxy for total metallicity: MgFe $\equiv \sqrt{\mbox{Mgb}\times\mbox{Fe4383}}$, where Mgb and Fe4383\footnote{This is the best combination of magnesium and iron indices we can use, since other indices observed by LEGA-C have typically larger errors, and are observed for a much lower number of galaxies (see Appendix A of \citealt{Bevacqua+23}.)} are provided by the DR3 of LEGA-C. In Appendix \ref{app:mgfe}, we show that MgFe correlates linearly with [M/H], as also predicted by models. Note that, although both indices are observed only for 183 galaxies in our sample \citep{Bevacqua+23}, this subsample covers the whole range of masses. 

To take into account the effect of kinematic broadening on the spectral indices, for each galaxy, we construct a model of the best-fitting age and metallicity by linearly interpolating the E-MILES models. Then, we broaden the model by convolving it with a gaussian of standard deviation corresponding to the galaxy’s velocity dispersion, $\sigma_*$ (as reported by the LEGA-C DR3). For each index, $I$, we thus calculate the correction factor $C = I_{\rm mod}/I_{\rm mod,broad}$, where $I_{\rm mod}$ is the model index measured at the LEGA-C spectral resolution, and $I_{\rm mod,broad}$ is the model index measured from the spectrum broadened by $\sigma_*$. We thus correct the observed indices, $I_{\rm obs}$, as $I_{\rm corr} = C \times I_{\rm obs}$.

In Fig. \ref{fig:meme_mgfe} we show the MgFe$-$mass diagram. The overall distribution of galaxies in Fig. \ref{fig:meme_mgfe} resembles that of Fig. \ref{fig:meme}. Indeed, while at low masses galaxies span a large range of MgFe values, at high masses galaxies have high MgFe. As highlighted by the dashed magenta line, the MEME relation is present even when using spectral indices. Interestingly, lower-mass galaxies (down to $\log_{10}$(M$_*$/M$_\odot) \sim 10.7$) reach MgFe values as high as galaxies $\sim1$ dex more massive. Crucially, MgFe does not depend on any model or fit.

\begin{figure*}\label{fig:sfh_ex}
\includegraphics[width=\textwidth]{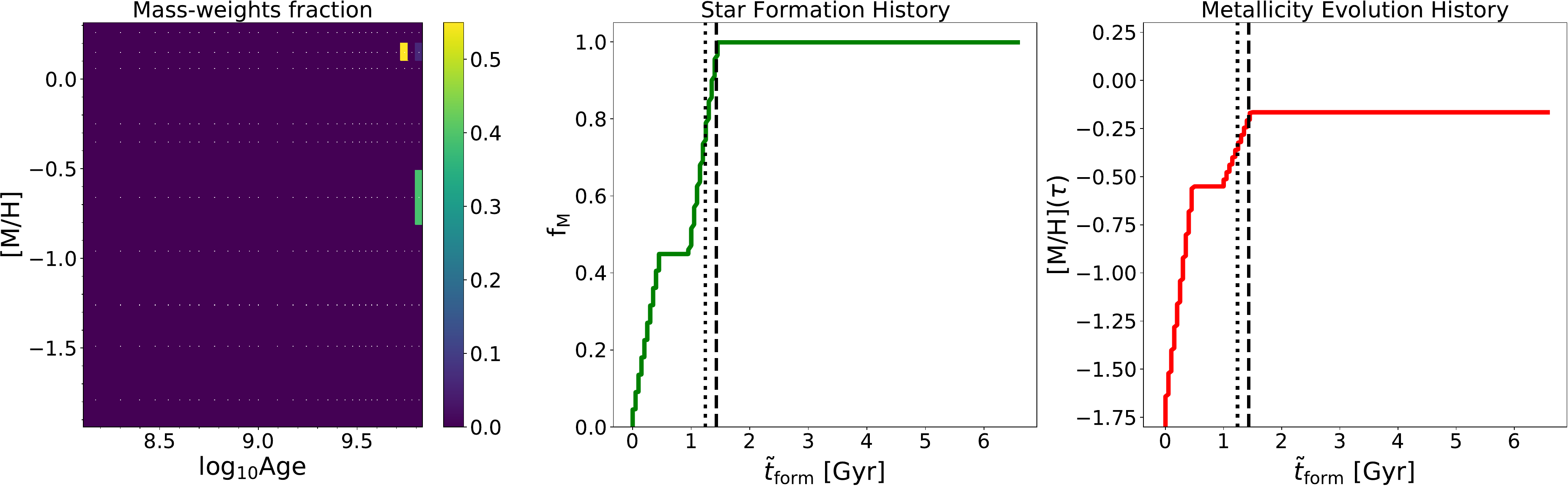}
\caption{Example of SFH and MEH for galaxy with LEGA-C ID: M23 - 151373. \textit{Left panel:} mass-weights map of the best fitting spectrum. The white dots in the map correspond to the combination of age and metallicity of each template used for the fits. The colorbar represents the mass weights assigned to SSPs in the best fitting spectrum. \textit{Middle panel:} SFH (solid green line) of the fitted galaxy, calculated as the cumulative sum of the rebinned mass weights, as a function of the cosmic formation time. \textit{Right panel:} MEH (solid red line) of the fitted galaxy, calculated using equation \eqref{eq:meh}, as a function of the cosmic formation time. In the middle and right panels, the dash and dotted lines indicates $\tau_{75}$ and $\tau_{90}$, namely the formation time at which the galaxy has reached 75$\%$ and $90\%$ of its mass, respectively.}
\label{fig:sfh_ex}
\end{figure*}

Results from both spectral fits and spectral indices indicate that the metallicity of quiescent galaxies is bounded by a mass-dependent lower limit that increases with stellar mass. We argue that the MZR is an empirical consequence of this lower limit, i.e. to the lack of massive, and metal-poor galaxies. Further, the MgFe index shows that low-mass galaxies can have metallicities as high as the most massive galaxies.

In Appendix \ref{app:simulations}, we show that simulations from both hydrodynamical and semi-analytical models predict a similar behavior to observations, with lower-mass galaxies spanning a larger metallicity range and reaching metallicities as high as the most massive galaxies, and exhibiting a mass-dependent lower limit. In particular, the predicted MEME relation is flatter than the observed one (see Fig. \ref{fig:tng_legac} and Fig. \ref{fig:gaea}). These predictions indicate that our results on LEGA-C galaxies are not a consequence of the methods used to measure their properties, or the S/N, but they are related to the physical processes driving the mass and metallicity of quiescent galaxies. We caution that the results shown in Appendix \ref{app:simulations} are qualitative. A proper comparison with observations would require careful and quantitative analysis of the models' data. However, this is beyond the scope of this work.

\section{The role of cosmic evolution on metallicity}\label{sect:cosmicevol}

The larger metallicity range spanned by lower mass galaxies could be a consequence of cosmic evolution affecting more significantly these galaxies than the most massive ones. In particular, it is known that, on average, lower mass galaxies take longer to form and to assemble their stellar mass (e.g., \citealt{McDermid+15}). A prolonged SFH can then affect the stellar population properties of a galaxy, thus varying its metallicity. Depending on the mechanisms increasing the stellar mass of galaxies (prolonged SF, mergers, gas accretion, etc.), the metallicity can in principle both increase or decrease. Indeed, quiescent galaxies exhibit a rich variety of SFHs \citep{DeLucia+06, Tacchella+22}. Therefore, the differential scatter in metallicity at different masses could be a consequence of both longer time scales of the mass assembly, and the variety of SFHs.

Another cosmic effect that can affect the general metallicity distribution is the progenitor bias \citep{progbias, Carollo+13, Poggianti+13}. Indeed,  the addition of newly quenched galaxies to the quiescent population can have different stellar population properties. 

If the observed metallicity distribution is a consequence of the cosmic evolution effects, we may expect that galaxies formed with similar SFHs or those that formed at earliest cosmic epochs follow a tighter MZR, i.e. galaxies should span a shorter range of metallicities. For these reasons, in this section we study whether the different SFHs or formation times can account for the differential metallicity scatter at different masses.

To this purpose, in section \ref{sect:SFH} we construct the SFHs of our galaxy sample and present our method to study the evolution of the metallicity of galaxies during their mass assembly process. Then, in section \ref{sect:met_tform} we study the dependence of the metallicity on different formation epochs, and how it affects the metallicity-mass diagram.

\begin{figure*}
\includegraphics[width=\textwidth]{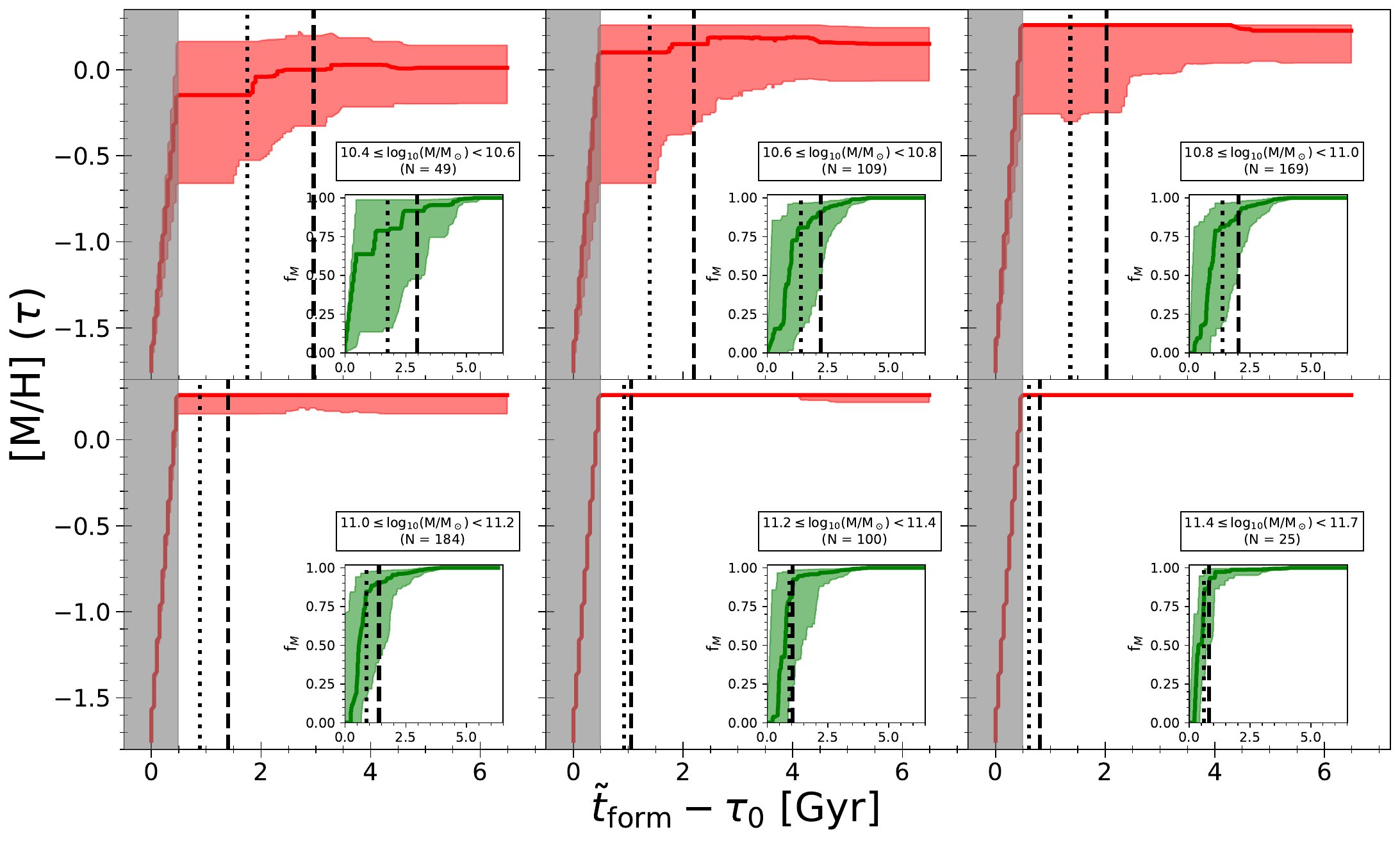}
\caption{MEHs (red) and SFHs (green, insight) for the whole sample of quiescent galaxies, divided into mass bins. In each panel a box indicating the number of galaxies, N, and the mass bin considered is shown. The solid red and green curves are the median MEH and SFH, respectively, calculated at each temporal step. The red and green shaded regions represent the 16th-84th percentiles. MEH curves show, by construction, a fictitious raise within the first 0.5 Gyr; for this reason, we shade this temporal region in grey. The black dotted and dashed lines are $\tau_{75}$ and $\tau_{90}$, respectively, corresponding to the time at which the median SFHs reach the 75$\%$ and 90$\%$ of their mass.}
\label{fig:MEH}
\end{figure*}

\subsection{Star Formation History and Metallicity Enrichment History}\label{sect:SFH}

\subsubsection{Methods}\label{sect:sfh_methods}

To track the mass assembly history of galaxies, we consider the mass-weights assigned by \texttt{pPXF} to the fitted SSPs, constituting the best-fitting spectrum. The SFH of a galaxy is calculated as the cumulative distribution function, $f_{\rm M}$, of the mass-weights as a function of the cosmic time. To have smoother curves, we perform a resampling of the SSPs by evenly redistributing weights into 10 bins between SSPs of adjacent ages. In the middle panel of Fig. \ref{fig:sfh_ex} we show an example of SFH relative to the best-fitting spectrum whose weights map is shown in the left panel. The SFH is plotted as a function of $\tilde{t}_{\rm form}$, defined as the formation time of the (rebinned) SSPs (i.e. it is calculated using the ages of the input SSPs instead of the average age of the galaxy, in equation \eqref{eq:tform}). We then define $\tau_x$ as the time at which a galaxy has reached $x$ percent of the total mass. In Fig. \ref{fig:sfh_ex} we show the times at which the considered galaxy has reached $75 \%$, and $90\%$ of its total mass, namely, $\tau_{75}$ and $\tau_{90}$, respectively.

The weights assigned by the fit to the input SSPs also provide information on how the metallicity evolves during the mass assembly. For instance, in the example shown in Fig. \ref{fig:sfh_ex}, the average metallicity is a combination of an older SSP with lower metallicity and a younger and more metallic SSP. This implies that, according to the fit, the metallicity of this galaxy has increased in time during the mass assembly process. Hence, we can reconstruct the Metallicity Enrichment History (MEH) of galaxies, and study how their metallicity has changed over time. 

Similarly to SFH, to build the MEHs we consider the mass-weights associated with each SSP, and evenly redistribute them into 10 bins to have smoother curves. Then, we sum the weighted metallicities and obtain the MEHs as a function of $t_{\rm form}$. Crucially, to perform this sum one can not simply use the cumulative function, as for the SFH. Indeed, the weights assigned by the fit represent the fractional contribution of each SSP to the total mass, i.e. to the already assembled galaxy. Instead, we aim to track the relative change of the metallicity following the mass assembly history\footnote{For instance, in the example shown in Fig. \ref{fig:sfh_ex}, within the first $\sim 2$ Gyr the only contribution to the global metallicity of the galaxy was due to the oldest SSP, having [M/H] = $-0.66$ dex. Then, the galaxy increased its stellar mass, and the fractional contribution of each SSP changes the global metallicity, reaching [M/H] = $-0.21$ dex. If we calculated the initial metallicity using the weights assigned by the fit, we would instead measure [M/H] $= 0.45 \times (-0.66) = -0.30$ dex (> -0.66 dex), thus minimizing the effective change in metallicity.}.

In terms of the fit, the metallicity, [M/H]($\tau$), reached by a galaxy at the time, $\tau$, when it has assembled a certain fraction of its total mass, depends on the weights assigned to the SSPs older than, or coeval to $\tau$. To keep track of the relative variation of metallicity with time we thus `update' the mass-weighted metallicity based on the fractional contribution of each SSP, from the oldest to the youngest. Numerically, [M/H]($\tau$) is calculated using the following equation:

\begin{equation}\label{eq:meh}
\mbox{[M/H]}(\tau) =\frac{\sum_{i}{ w_i (\tilde{t}_{\mbox{\scriptsize{form}} ,i} \leq \tau) \times \mbox{[M/H]}_i (\tilde{t}_{\mbox{\scriptsize{form}} ,i} \leq \tau)}}{\sum_{i}{w_i(\tilde{t}_{\mbox{\scriptsize{form}} ,i} \leq \tau)}}
\end{equation}

where the sums are performed over the (rebinned) SSPs, with ages corresponding to $\tilde{t}_{\rm form} \leq \tau$, i.e. with ages older than the youngest SSP in the temporal bin considered. In the right panel of Fig. \ref{fig:sfh_ex} we show an example of MEH. 

\subsubsection{The dependence of SFHs and MEHs on stellar mass}

\begin{figure}
\includegraphics[width=\columnwidth]{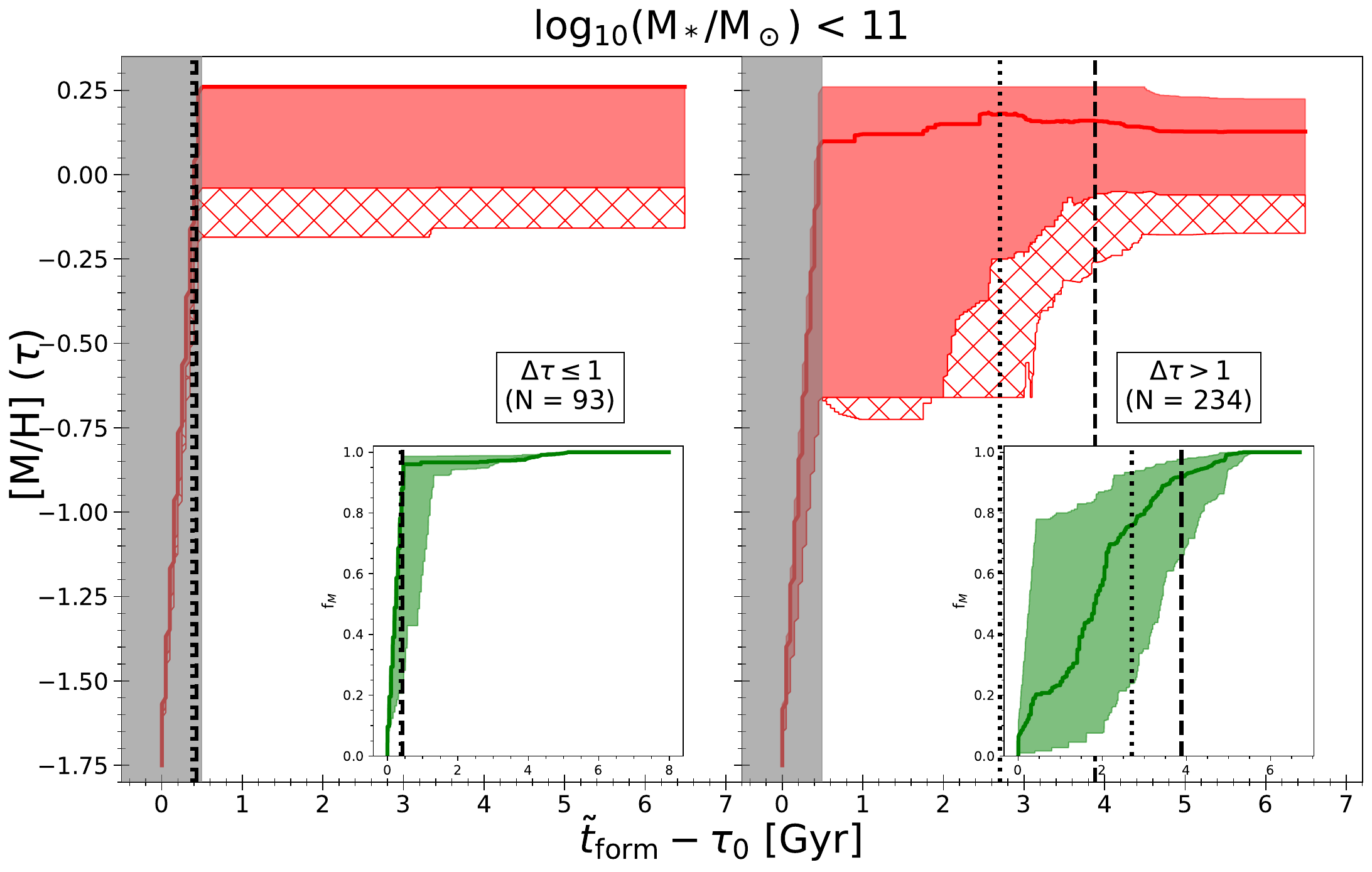}
\caption{MEHs and SFHs of all galaxies with log$_{10}$(M$_*$/M$_\odot$) < 11 together, divided according to their SFH, between galaxies that formed within $\Delta\tau = \tau_{95} - \tau_{5} = $ lower and greater than 1 Gyr. Notice that, since we are here considering galaxies at all redshifts, given the mass completeness limit, statistics are impaired towards higher mass galaxies. The hatched region indicates the 16th percentile of galaxies in the lowest mass bin, $10.4 \leq \log_{10}$(M$_*$/M$_\odot) < 10.6$.}
\label{fig:MEH_lm}
\end{figure}

We build SFH and MEH curves for all quiescent galaxies in our sample and divide them into six mass bins at steps of 0.2 dex between $10.4 \leq \log_{10}$(M$_*$/M$_\odot) < 11.7$. In Fig. \ref{fig:MEH} we show the median MEH, and SFH curves for different mass bins, shaded between `envelopes' of the 16th and 84th percentiles. To produce these plots, all curves have been rescaled to the $\tilde{t}_{\rm form}$ of the oldest SSP with non-zero weight ($\tau_0$), in order to have the same zero point in time (virtually, when the galaxy started forming stars). 

In Appendix \ref{app:MEH} we discuss how our choices on the fitting method affect the average SFH and MEH curves, as well as single cases. Briefly, we find that fitting a larger wavelength range or including older SSPs in the fit extends the average duration of the SFH (with a smaller impact at increasingly higher mass), while MEHs have slightly larger percentiles' envelopes but without changing significantly the curves. Similarly, different S/N ratios can introduce some scatter around single curves, without affecting the trends. We conclude that the estimates of $\tau$ and [M/H]($\tau$) of single galaxies may depend on the fitting procedure, but the general trends do not.

Fig. \ref{fig:MEH} shows that the median MEH curves have higher metallicities as the stellar mass increases, reaching the maximum value allowed by models, [M/H] = $+0.26$, at $10.8 \leq \log_{10}$(M$_*$/M$_\odot) < 11$. The percentiles envelopes reduce at higher masses, and essentially coincide with the average trends for $\log_{10}$(M$_*$/M$_\odot) \geq 11$, indicating that the range of metallicities reduces as the mass increases, as pointed out in section \ref{sect:meme}. Compared to the SFHs, the MEHs indicate that, on average, in an extended SFH the stellar metallicity tends to increase during the star formation phase. However, this effect is mild. Indeed, within the time interval $\Delta\tau \equiv \tau_{95} - \tau_{5}$, we estimate a median increase of $\lesssim 0.2$ dex in the lowest mass bin, that reduces to zero at higher masses. This result is not surprising given that, at all mass bins, galaxies formed $\sim 75\%$ of their stellar mass within the first couple of Gyr\footnote{These values are also robust against the different fitting method considered in Appendix \ref{app:MEH}.}, on average, indicating that the greatest fraction of the stellar mass of quiescent galaxies at all masses formed quickly. This implies that subsequent star formation has a limited ($\leq 25\%$ in mass) impact on their final metallicity. 

We further investigate SFHs and MEHs of low and high mass galaxies, separately.

At high masses ($\geq 10^{11}$M$_\odot$), the median curves are well representative of the individual SFHs and MEHs. Indeed, most galaxies formed almost all of their mass very quickly ($\Delta\tau \sim 1$ Gyr), in agreement with several other studies (e.g., \citealt{Thomas+10, McDermid+15}). This implies that even though subsequent star formation occurs, the contribution of the newly formed stellar mass is too low ($\leq 10\%$) to have a relevant impact on the global metallicity, which remains essentially unchanged.

On the other hand, on average, lower-mass (M$_* < 10^{11}$M$_\odot$) galaxies take longer ($\Delta\tau \sim 2 - 3$ Gyr) to form most of their mass increasing their metallicity over the time of the star formation. However, the median trends are not fully representative of the rich variety of SFHs and MEHs exhibited by lower mass galaxies (in agreement with \citealt{Tacchella+22}). In fact, we find that $\sim 70\%$ of low-mass galaxies has an extended SFH, and in particular for $\sim 55\%$ of the cases the metallicity increases over this time, while for $\sim 15\%$ it decreases; the remaining $\sim 30\%$ of cases have already formed $\geq 90\%$ of their mass within $\sim$ 1 Gyr, similarly to more massive galaxies, thus leaving their metallicity essentially unchanged afterward\footnote{In appendix \ref{app:MEH} we show that these statistics are not affected by the S/N of the spectra.}.

For this reason, we further divide low-mass galaxies into two subsamples: those with $\Delta\tau \leq 1$ Gyr, and those with $\Delta\tau > 1$ Gyr, this time interval corresponding to the temporal resolution of the E-MILES models. We show the MEH and SFH for these two subsamples in Fig. \ref{fig:MEH_lm}. Note that we are here including galaxies at all redshift bins. Consequently, the curves are biased towards higher masses, given their larger statistics because of the mass completeness limit.

The metallicity range spanned by low-mass galaxies whose stellar mass formed in $\Delta\tau \leq 1$ Gyr is $\sim 0.3$ dex large, between the 16th and 84th percentiles (and up to $\lesssim 0.5$ dex when considering only galaxies with $10.4 \leq \log_{10}$(M$_*$/M$_\odot) < 10.6$). This is considerably larger than the range spanned by high-mass galaxies ($<<$ 0.1 dex) with similar SFH. This implies that the stellar mass of a galaxy plays a more important role than the `shape' of the SFH in determining the final metallicity. We note that high-mass galaxies with $\Delta\tau\leq1$ Gyr could intrinsically have a $\Delta\tau$ smaller than low-mass galaxies with $\Delta\tau\leq1$ Gyr. However, the age resolution of the E-MILES models does not allow us to establish if this is the case. Thus, we can not explore the effect of the SFH on shorter time scales. 

Lower-mass galaxies with $\Delta\tau > 1$ Gyr typically increase their metallicity during the mass assembly process. We estimate a median variation of $\approx +0.1$ dex over $\Delta\tau$. Moreover, the 16th - 84th percentiles for the final metallicities of galaxies that formed in $\Delta\tau > 1 $ Gyr is comparable to those that formed in $\Delta\tau \leq 1$ (by performing a KS test on the final metallicity of the two subsamples, we get a p-value of 0.39). This implies that it is not the variety of SFHs that determines the different metallicity values spanned by galaxies, at fixed stellar mass.

Finally, we point out that TNG50 simulations predict that higher-mass galaxies have higher fractions of ex-situ stars (see Appendix \ref{app:simulations}). Interestingly, lower-mass galaxies are mostly composed of in-situ stars, implying that the scatter in metallicity is not due to mergers.

To summarize, the MEHs of quiescent galaxies indicate that stellar mass plays a more important role than SFH in determining the final metallicity of galaxies. In most cases, metallicity increases along the SFH of a galaxy. The median increase is however mild.

We conclude that the observed metallicity distribution of quiescent galaxies is not a consequence of different time scales of the star formation, or of the variety of SFHs. The role of the SFH is secondary to that of the stellar mass. If SFH plays an important role in determining the metallicity, it is on time scales shorter than 1 Gyr.

\subsection{The evolution of the mass-metallicity diagram as a function of the formation time}\label{sect:met_tform}
\begin{figure}
\includegraphics[width=\columnwidth]{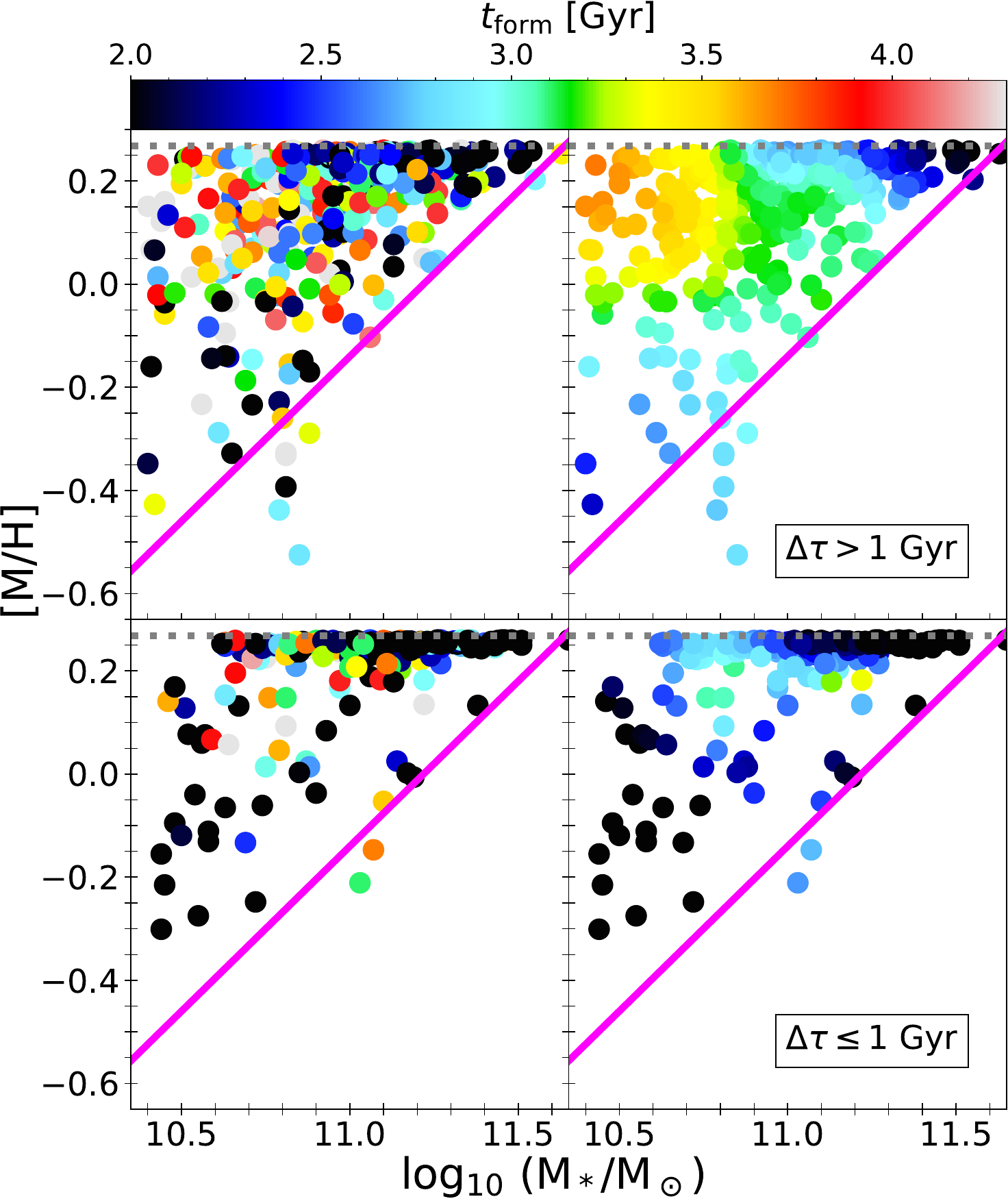}
\caption{Metallicity-Mass diagram colored with $t_{form}$. Upper panels show the subsample of galaxies with $\Delta\tau > 1$ Gyr, while lower panels those with $\Delta\tau \leq 1$ Gyr. On the left panels, colors correspond to the estimated values of each point; on the right panels, colors have been smoothed using \texttt{LOESS} to emphasize any possible trend with $t_{\rm form}$.}
\label{fig:meme_tform}
\end{figure}

In Fig. \ref{fig:meme_tform}, we show the metallicity - mass diagram color-coded by $t_{\rm form}$. Since it represents the average formation time of the stellar population, $t_{\rm form}$ is degenerate with $\Delta\tau$. Then, for galaxies that experienced a prolonged SFH, $t_{\rm form}$ is the result of a combination of the `initial' epoch of stellar formation, and subsequent stellar bursts or extended SF occurring at later times. On the other hand, for galaxies that formed the majority of their stellar mass within short times, $t_{\rm form}$ is in a sense better representative of the cosmic epoch at which galaxies formed because most of the stellar mass is coeval (within the 1 Gyr resolution limit of the models). For this reason, in Fig. \ref{fig:meme_tform} we split the sample into two: galaxies that formed with $\Delta\tau > 1$ Gyr, and within $\Delta\tau \leq 1$ Gyr.

In the upper panels of Fig. \ref{fig:meme_tform}, we show galaxies with $\Delta\tau > 1$. Galaxies whose average stellar mass formed at later cosmic times typically have lower mass and higher metallicity. This is in agreement with the results shown by MEHs since the largest fraction of low-mass galaxies has an extended SFH (thus resulting in later $t_{\rm form}$), typically increasing the metallicity of galaxies over time. The observed trend could be due to the combination of the extension of the SFH, and the progenitor bias \citep{Belli+19}. However, it is not possible to disentangle the two effects here.

\begin{figure}
\includegraphics[width=\columnwidth]{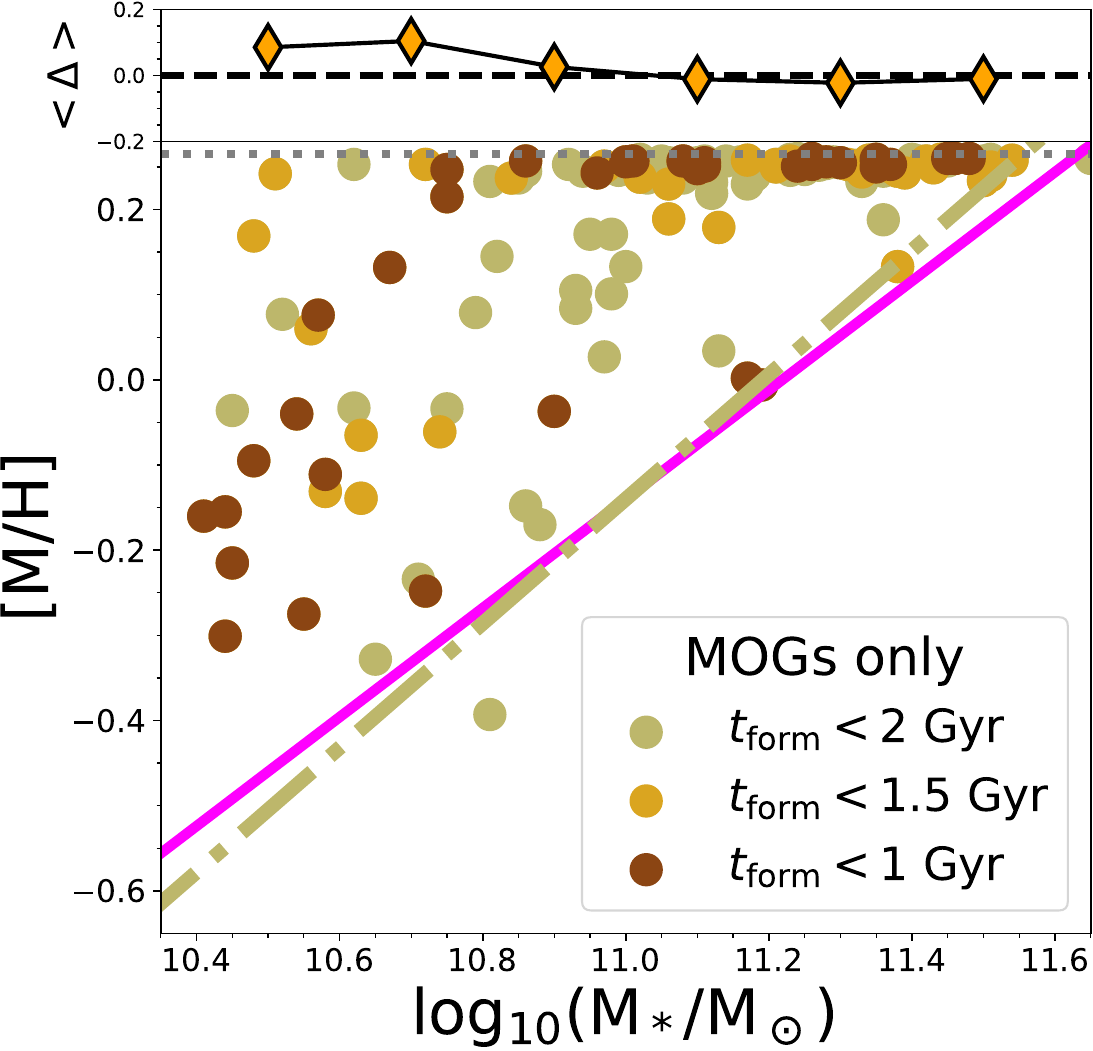}
\caption{Metallicity-Mass diagram of MOGs. Different colors correspond to different definitions of MOGs. The magenta line is the MEME relation for the whole sample, while the dashdotted khaki line is the MEME for MOGs with $t_{\rm form} < 2$ Gyr. The upper panel shows the difference between the average metallicity values of the full sample and of MOGs, $<\Delta>$, at different mass bins.}
\label{fig:meme_mog}
\end{figure}

\begin{figure}
\includegraphics[width=\columnwidth]{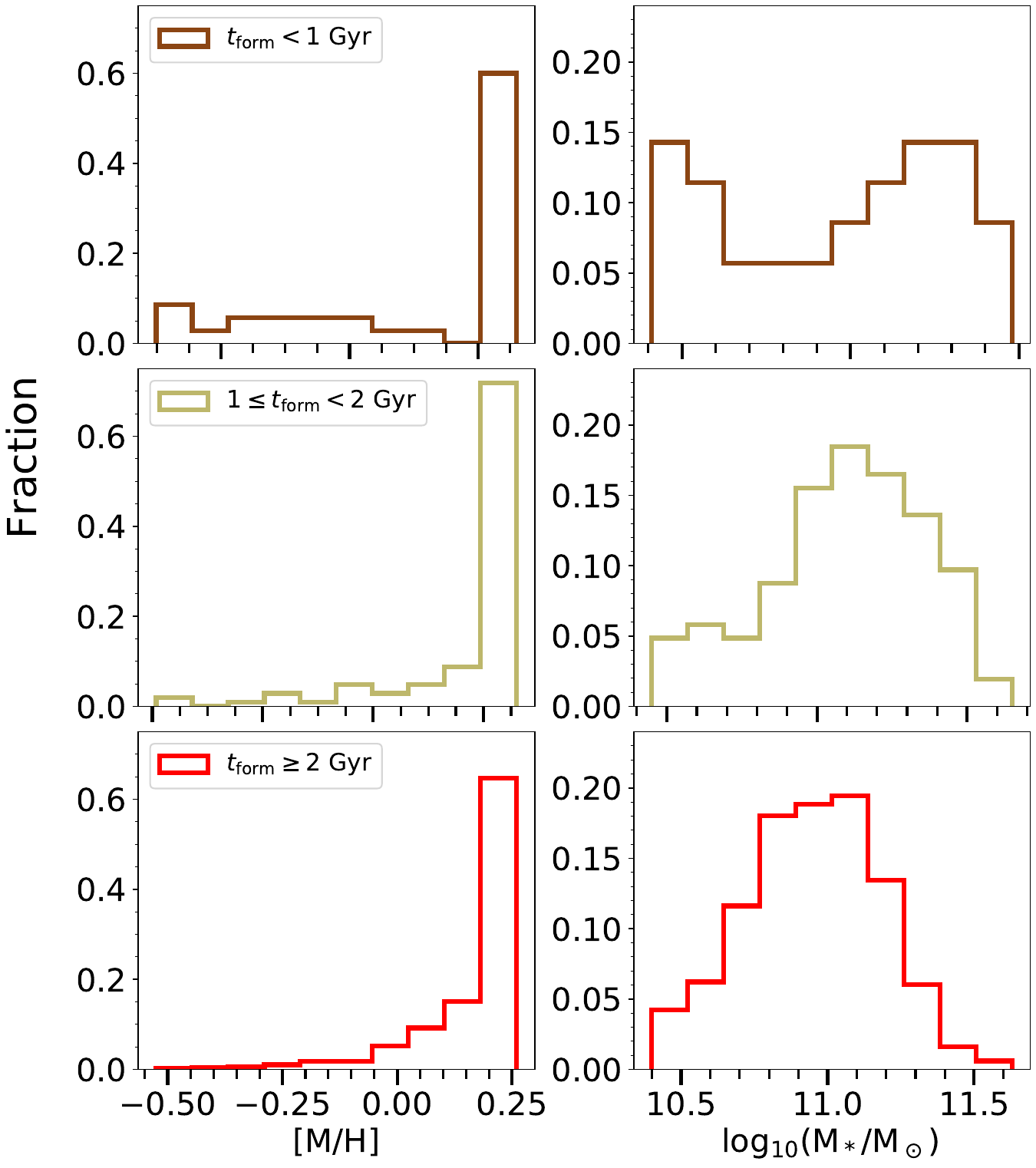}
\caption{Histograms of stellar metallicity (left panels) and mass (right panels) for galaxies formed at $t_{\rm form} < 1$ Gyr (top panels), $1 \leq t_{\rm form} < 2$ Gyr (middle panels), and $t_{\rm form} \geq 2$ Gyr (bottom panels).}
\label{fig:mogs_hist}
\end{figure}

In the lower panels of Fig. \ref{fig:meme_tform}, we show galaxies with $\Delta\tau \leq 1$. The general trend with $t_{\rm form}$ is similar to the one of galaxies with $\Delta\tau > 1$. Here, the degeneracy of $\Delta\tau$ with $t_{\rm form}$ is minimized, since we are considering very short SFHs. This might suggest that galaxies formed at later cosmic times tend to be more metallic, as also found, e.g., by \citealt{Saracco+23}, regardless of the SFH. However, we did not find any statistically significant correlation between $t_{\rm form}$ and M$_*$, as the scatter in our data is too large and the relation with [M/H] is hampered by the limitation of the models.

We also note that TNG50 simulations indicate that the greatest fraction of the stellar mass of lower-mass galaxies formed in situ (see Fig. \ref{fig:tng_legac}). Thus, according to simulations, the trend with $t_{\rm form}$ at lower masses is not affected by mergers.

To further probe the effect of the formation time, we consider the extreme cases of Maximally Old Galaxies (MOGs), defined as those galaxies with $t_{\rm form} < 2$ Gyr\footnote{Although the choice of 2 Gyr is arbitrary, our results do not change when considering more stringent constraints, while significantly lowering the statistics (see Fig. \ref{fig:meme_mog}). We further checked that the age-sensitive indices (H$\beta$, H$\delta$, H$\gamma$) of these galaxies indicate that they are generally old.}. By definition, MOGs host stellar populations that formed at earliest cosmic times, regardless of the redshift at which they are observed, or their SFH. The fact that we can measure such old ages implies that these galaxies have not experienced any major SF event after 2 Gyr ($z \sim 3$) from the Big Bang meaning that they evolved passively for the great majority of their life, or have been subject to cosmological events that have not affected significantly their average stellar population properties.  

In Fig. \ref{fig:meme_mog} we show the metallicity-mass diagram of MOGs. As evident, MOGs distribute all over the metallicity - mass plane. Interestingly, the MEME relation estimated for MOGs (as in section \ref{sect:meme}) is essentially the same as the one found for the whole sample. Further, the maximum metallicity is reached for both lower- and higher-mass galaxies. The plot indicates that the MZR of MOGs was already in place at $z \sim 3$. In the upper panel of Fig. \ref{fig:meme_mog} we show the difference, $<\Delta>$, between the average metallicity of the full sample and of MOGs, at different mass bins. The plot indicates that the MZR was mildly lower ($<\Delta> \sim 0.1$ dex) at low ($\leq10^6$M$_\odot$) stellar masses, while it did not change at higher masses ($<\Delta> \sim 0.0$ dex), since $z \sim 3$. We point out that the same dependence of the metallicity on the mass for MOGs has also been found by \cite{Saracco+23}, who studied a sample of quiescent galaxies at redshift $z \sim 1.1$, using a different analysis.

By performing a KS test comparing the metallicity of MOGs with that of the whole sample (including MOGs) we find a very low p-value ($\sim 10^{-4}$), indicating that the metallicity distribution of MOGs is not representative of the total sample of quiescent galaxies. In particular, by comparing MOGs with non-MOGs we get a p-value of $\sim 10^{-7}$, implying that the metallicity distribution changes at different cosmic times. The same result is also found when comparing the metallicity distributions of extreme MOGs, having $t_{\rm form} < 1$ Gyr, with the remaining sample of MOGs, implying a differential evolution of the metallicity - mass diagram even over temporal scales of the order of 1 Gyr, and at earliest cosmic times. 

These results indicate that galaxies distribute differently on the metallicity - mass plane as a function of the cosmic time at which they formed. This is further illustrated in Fig. \ref{fig:mogs_hist}. This figure shows that, at later cosmic times, the relative number of galaxies with very low metallicities decreases. On the other hand, the mass distribution change from a bimodal distribution at earliest cosmic times, where low mass galaxies have low metallicities and massive galaxies are metal-rich, towards a unimodal distribution peaked at intermediate ($\log_{10}$(M$_*$/M$_\odot) \sim 11$) masses. 


\section{Summary}\label{sect:summary}

In this paper, we have studied the relation between the stellar mass and metallicity for a sample of 637 quiescent galaxies selected from the LEGA-C survey at $0.6 \leq z \leq 1$, with stellar masses $10.4 \leq \log_{10}($M$_*$/M$_\odot) < 11.7$. Stellar metallicity, star formation history, and metallicity evolution history have been derived through full spectral fitting with E-MILES models \citep{EMILES}. We summarize our results as follows:

\begin{enumerate}[1.]

\item While massive galaxies always have high, supersolar metallicities, low-mass galaxies can be both metal-rich and -poor, spanning the range $-0.6 \lesssim$[M/H]$\leq 0.26$ dex (see Fig. \ref{fig:meme}). This systematic empirically defines a MEtallicity-Mass Exclusion (MEME) zone in the metallicity - mass diagram, delimited by a mass-dependent lower boundary to metallicity, that we have defined as the MEME relation (eq. \eqref{eq:meme}).

\item Using a combined index MgFe, as a proxy for the metallicity, we recover the MEME relation (Fig. \ref{fig:meme_mgfe}), implying that the existence of this limit is not an artifact of either the full spectral fitting or the models. Further, lower-mass galaxies reach MgFe values as high as galaxies almost $\sim 1$ dex more massive mass.


\item To evaluate how the metallicity of galaxies changes during star formation, we constructed the Metallicity Enrichment Histories (Fig. \ref{fig:MEH}). We found that:
\begin{itemize}

\item Massive ($\geq 10^{11}$M$_\odot$) galaxies formed the greatest fraction of their stellar mass ($\geq 90\%$) very quickly ($\Delta\tau \leq 1$ Gyr), with median metallicities reaching the highest value allowed by models (0.26 dex). Therefore, any subsequent SF activity has a negligible ($\leq 10 \%$) impact on the global metallicity. 

\item Lower-mass ($< 10^{11}$M$_\odot$) galaxies, show a rich variety of SFHs. About 1/3 of these galaxies formed most of their mass very quickly ($\Delta\tau \leq 1$ Gyr), similar to massive galaxies. However, the metallicity range spanned by the massive galaxies ($<<0.1$dex) is much lower than that spanned by lower mass galaxies ($\sim 0.3 - 0.5$ dex) with similar SFH. The remaining 2/3 of low-mass galaxies have an extended SFH. For the great majority of these cases, the metallicity increases during the SF (while for a small fraction, 15$\%$, the metallicity decreases) by $\sim 0.1$ dex, on average. Finally, we find that galaxies with very different SFH span similar metallicity ranges.

\end{itemize}

\item We evaluated how the metallicity - mass diagram changes as a function of the formation time (Fig. \ref{fig:meme_tform}). We found that galaxies formed at later cosmic times ($t_{\rm form} > 2$ Gyr) typically have low/intermediate masses ($\log_{10}$(M$_*$/M$_\odot) \leq 11$) and high (supersolar) metallicities. This could be due to the combination of extended SFH and later formation time.

\item We found that Maximally Old Galaxies (MOGs, i.e. galaxies with $t_{\rm form} \leq 2$ Gyr), for which the effect of SFH is minimized, also distribute differently on the metallicity-mass diagram at different cosmic times. In particular, extreme MOGs ($t_{\rm form} < 1$ Gyr) have low metallicities at low masses, and high metallicities at high masses, while galaxies that formed later have intermediate masses and higher metallicities. Interestingly, MOGs are distributed all over the metallicity-mass plane, implying that the upper bound and the MEME relation, and hence the MZR are already in place at $z>3$. We estimated a mild ($<\Delta> \sim 0.1$ dex) evolution of the MZR at low ($\log_{10}$(M$_*$/M$_\odot) \leq 10.6$) masses, while no significant evolution is found at higher masses.

\end{enumerate}

We discuss these results in the following section.

\section{Discussion}\label{sect:discussion}

Before discussing our results, it is important to keep in mind that in this work we have considered galaxies with $\log_{10}$(M$_*$/M$_\odot) \geq 10.4$. Indeed, there is evidence for a break in the MZR relation at lower masses (e.g., \citealt{Blanc+19}), where the slope of the MZR becomes steeper \citep{Gallazzi+05,Panter+08, Sybilska+17,Blanc+19, Zhuang+21}. Therefore, our considerations are limited to the high-mass end of galaxies.

Numerous studies have already investigated the relation between metallicity and mass of galaxies \citep{Tremonti+04, Thomas+2005, Gallazzi+05,  Panter+08, Thomas+10, Gallazzi+14, Choi+14, McDermid+15, Ginolfi+20}. The conclusions are often based on the average trend, i.e. the MZR, suggesting that more massive galaxies are more metal-rich than the less massive ones.

In this work, we have put emphasis on the fact that while massive galaxies are indeed metal-rich, low-mass galaxies can have both low metallicities, and metallicities as high as the most massive ones over $\sim1$ dex in stellar mass. Therefore, the MZR is not fully representative of the general metallicity distribution. Indeed, we note that the average increase in metallicity with mass is due to the lack of massive and metal-poor galaxies, rather than to lower-mass galaxies having lower metallicities.

We showed that the metallicity of quiescent galaxies is bounded by a mass-dependent lower limit, i.e. the MEME relation, which has been independently recovered from full-spectral fitting and from the spectral index MgFe, and is also predicted by simulations (see Appendix \ref{app:simulations}). The existence of a lower limit is however difficult to explain. The presence of the MEME relation for galaxies whose stellar mass formed at $z\geq3$ suggests that this lower limit has been set at high redshifts, and it could be linked to the physical conditions of galaxy formation at the earliest cosmic times.

The fact that lower-mass galaxies can reach metallicities as high as the most massive galaxies, as seen in the MgFe index and in simulations, has been already pointed out in other studies about the gas-phase metallicity of star-forming galaxies (e.g., \citealt{Tremonti+04, Mannucci+10}). This might hint to the existence of an upper limit to metallicity that could be explained by the inability of stars to produce more metals than a certain fraction (e.g., \citealt{Parmentier+99}), regardless of the condition at the formation, and the ability of a galaxy to retain metals. The empirical lowering of the maximum metallicity reached by lower-mass galaxies, observed at $\log_{10}($M$_*$/M$_\odot) < 10.4$ in other works (e.g. \citealt{Blanc+19}), could then be a combined effect of the inability of both producing and retaining the largest fraction of metals allowed by the stellar physics.


In general, the effect of the cosmic evolution is to populate differently the metallicity-mass diagram, within these upper and lower boundaries. Indeed, most of the quiescent galaxies in our sample exhibiting an extended SFH, and those that formed at lower redshifts have low masses ($\leq 10^{11}$M$_\odot$) \citep{McDermid+15}. This indicates that the metallicity of lower-mass galaxies is more easily affected by later star formation, and that the progenitor bias could affect the metallicity distribution at low masses. Instead, simulations indicate that lower-mass galaxies are not significantly affected by mergers, being composed mostly of stars formed in situ. Most massive galaxies, on the other hand, formed almost all their stellar mass quickly, and at early cosmic times, in agreement with previous studies, and with the downsizing \citep{Cowie+96, Thomas+10}. Their metallicity has not changed significantly since their formation. 

Further, galaxies on the metallicity-mass diagram distribute differently at different cosmic times, having low/intermediate masses and high (supersolar) metallicities at increasing $t_{\rm form}$. These results suggest a mass-dependent evolution of the average metallicity, as shown in Fig. \ref{fig:meme_mog}, in agreement with some previous observational works (e.g., \citealt{Gallazzi+14}), and simulations (e.g., \citealt{DeRossi+17}), rather than a rigid shift of the MZR (as found, e.g., by \citealt{Leethochawalit+18}). 

The fact that star-formation in high-mass galaxies took place within short time-scales, in a deep potential well, at early epochs, might explain the absence of massive, metal-poor galaxies. A short and early starburst implies a high star formation efficiency, indicating high levels of dissipation at high redshift and an efficient enrichment of the ISM with metals that are retained by the large potential well, thus resulting in an old and metal-enhanced massive, compact galaxy (e.g., \citealt{Khochfar+2006}). For the same short formation time, we can suppose that the metallicity decreases as the mass decreases because of the lower ability to retain metals, as well as a generally lower star formation efficiency. As a consequence, low-mass galaxies with $t_{\rm form} < 1$ Gyr have the lowest metallicity. To reach metallicity as high as the metallicity of massive galaxies longer star formation, or later formation times are needed to enrich the ISM.




We conclude by pointing out that investigating in more details the overall metallicity distribution, instead of just the average trend, is valuable to put stringent constraints on the complex relation between the mass and the metallicity of galaxies. Observationally, it can alleviate the tension between different methods used to estimate the metallicity, and between the different, interdependent proxies of the galaxy's mass that often lead to different results (e.g., \citealt{Barone+20, Baker+23}). Further, investigating the metallicity limits using simulations and metallicity evolutionary models (e.g., \citealt{Peng+15}) could provide stringent constraints on the physical processes determining the final metallicity of galaxies.

\begin{acknowledgements}
We thank the anonymous referee for useful comments that helped improving this paper. D.B., P.S., R.D.P.,  F.L.B., A.P., and C.S. acknowledge support by the grant PRIN-INAF-2019 1.05.01.85.  C.T. acknowledges the INAF grant 2022 LEMON. G.D. acknowledges support by UKRI-STFC grants: ST/T003081/1 and ST/X001857/1
\end{acknowledgements}

%
%

\bibliographystyle{aa} 
\bibliography{biblio1.bib} 

\begin{appendix}
\section{Tests on spectral fitting}\label{app:fittest}

\subsection{Fitting with older input SSP}\label{app:age_ssp}

As discussed in section \ref{sect:methods}, we impose a tight constraint on the upper limit of the input ages of SSPs, corresponding to a maximum input age of 6.5 Gyr for the E-MILES models. To evaluate how our age and metallicity estimates depend on such constraint, here we perform fits allowing older input SSPs. To this aim, we consider a random sample of 50 galaxies in the lowest redshift bin ($0.6 \leq z \leq 0.7$). We ensure that the subsample is well representative of the whole sample since it spans a wide range of masses, S/N, ages, and metallicities.

To perform the fits, we use the same method described in section \ref{sect:methods}. We set three different upper limits to the age of the input SSPs: 7 Gyr, 8 Gyr, and 10 Gyr. The first two limits are equivalent to ignoring our assumption that galaxies started forming stars at $z=10$, or/and considering the age of the Universe at the lowest redshift of LEGA-C galaxies, $z=0.6$ ($\approx 7.8$ Gyr). The 10 Gyr limit, albeit physically incorrect, allows us to evaluate a possible degeneracy between age and metallicity affecting the estimated properties.

\begin{figure}
\includegraphics[width=\columnwidth]{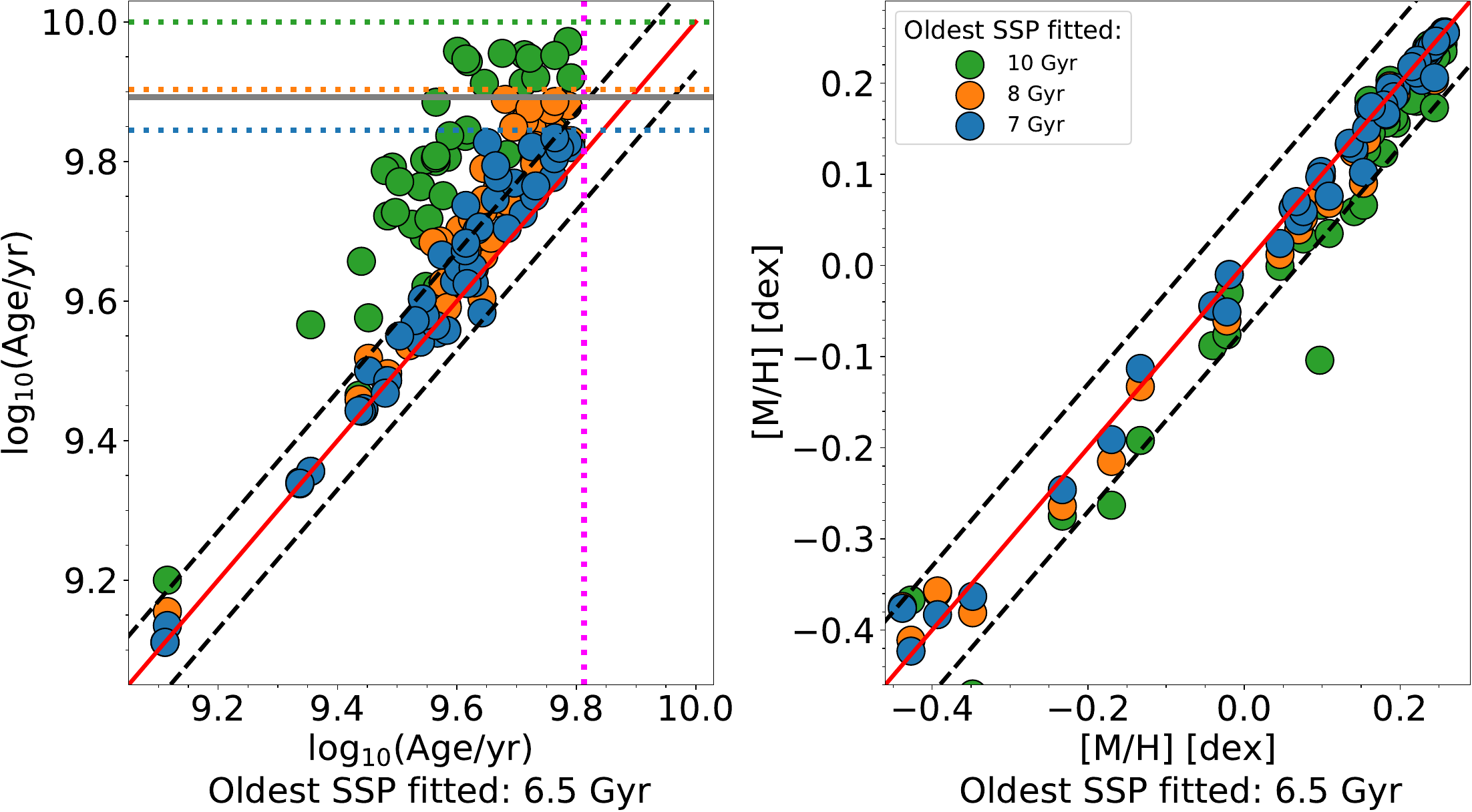}
\caption{Comparison between estimated ages (left panel) and metallicities (right panel) using different upper limits on the age of the input SSPs. Blue, orange, and green dots represents the estimated ages using a 7, 8, and 10 Gyr age limit, respectively, marked in the left panel by the corresponding horizontal dotted lines. The magenta vertical line corresponds to the 6.5 Gyr age limit imposed in section \ref{sect:methods}. The grey horizontal line is the age of the Universe at $z=0.6$ ($\approx 7.8$ Gyr). On both panels, the red line is the one-to-one relation, while the black dashed lines are the 1$\sigma_{\scriptsize{\rm Age}}$ and 1$\sigma_{\scriptsize{\rm [M/H]}}$ uncertainties.}
\label{fig:app_ssp}
\end{figure}

In Fig. \ref{fig:app_ssp} we compare these estimates with those estimated using a maximum age limit of 6.5 Gyr. Considering the fits that include the 7 Gyr old SSPs, fitting older SSPs templates provides ages biased towards older values by $\sim0.04$ dex, on average, although ages are in most cases consistent within 1$\sigma_{\rm Age}$ = 0.07 dex. With the 8 Gyr limit, the average offset is 0.07 dex, i.e. equal to $\sigma_{\rm Age}$. On the other hand, metallicities are generally lower, but the bias is very small, $\sim 0.01$ and $\sim 0.03$ dex, on average, for the 7 and 8 Gyr limits, respectively. Thus, metallicity estimates are largely consistent within the 1$\sigma_{\rm [M/H]}$ = 0.06 dex. Finally, fitting up to 10 Gyr provides ages significantly older (up to 0.3 dex), and many galaxies have ages much older than the age of the Universe. Nevertheless, metallicities are still pretty much consistent within the 1$\sigma_{\rm [M/H]}$ = 0.06 dex (they are 0.05 dex lower, on average).

Considering the age estimates with the 7 (8) Gyr limit, 10$\%$ (20$\%$) of galaxies deviate significantly ($>2\sigma_{\rm Age}$) from the one-to-one relation. We find that the estimated metallicities for almost all these galaxies are close to the maximum value allowed by models ($0.26$ dex). When fitting these spectra with SSPs up to 10 Gyr old, we obtain ages older than (or comparable to) the age of the Universe at $z=0.6$ ($\approx 7.8$ Gyr). Since such ages are unphysical, these galaxies likely have higher metallicities than the maximum metallicity of models, and the fit provides older ages to `compensate' for the lack of higher metallicity models. We find no clear dependence on stellar mass or S/N for these galaxies.

Overall, fitting older SSPs (up to 7 and 8 Gyr) typically provides older ages, but in most cases, they are consistent within the typical uncertainty. The most deviating age estimates are likely due to the age-metallicity degeneracy. In any case, the impact on metallicity is negligible, and all estimates are consistent within the errors.

\subsection{Multiplicative polynomials vs. reddening curve}\label{app:mdeg_redd}

\begin{figure}
     \centering
     \begin{subfigure}[b]{0.49\columnwidth}
         \centering
         \includegraphics[width=\textwidth]{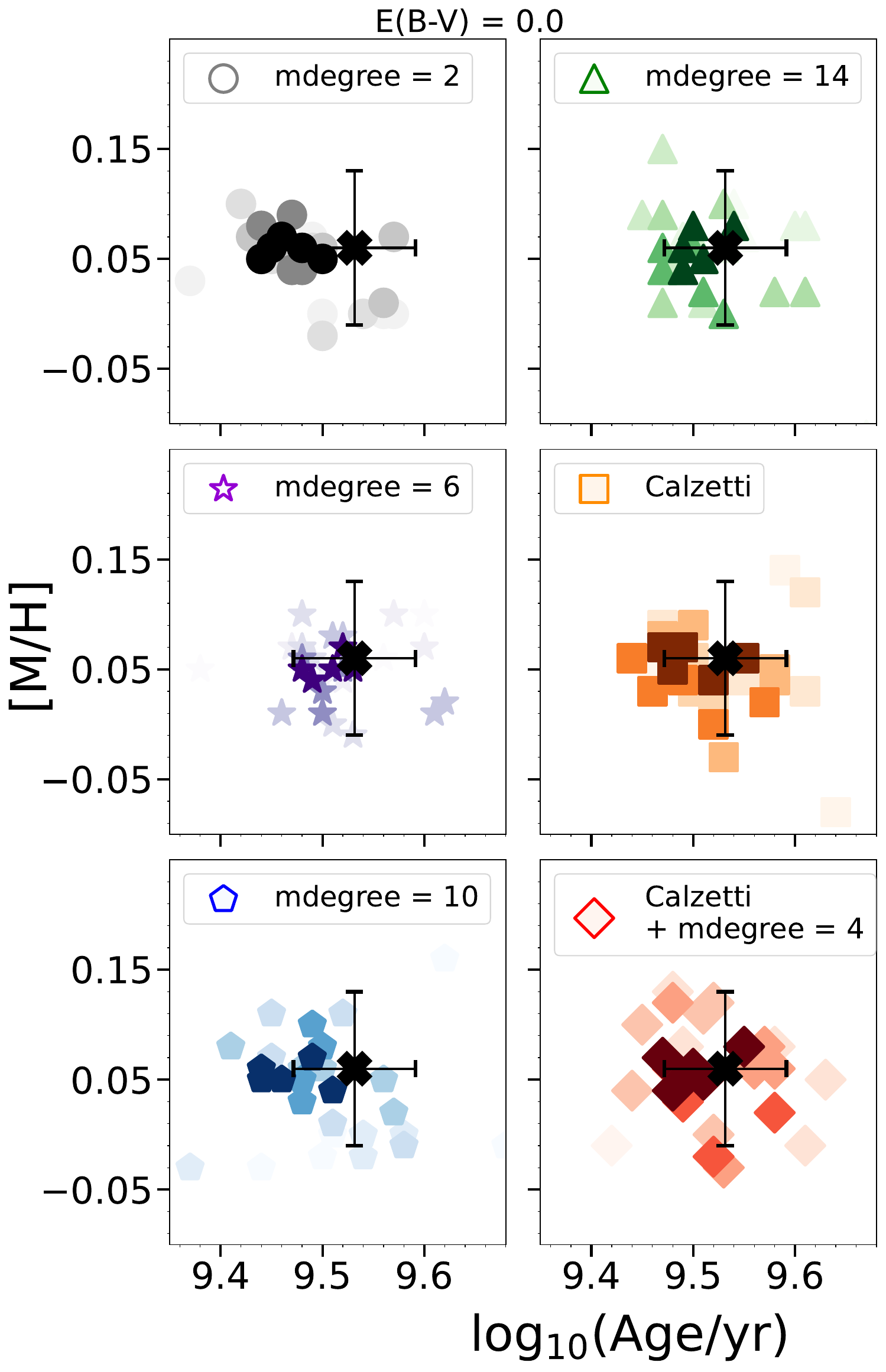}
         \caption{}
     \end{subfigure}
     \hfill
     \begin{subfigure}[b]{0.49\columnwidth}
         \centering
         \includegraphics[width=\textwidth]{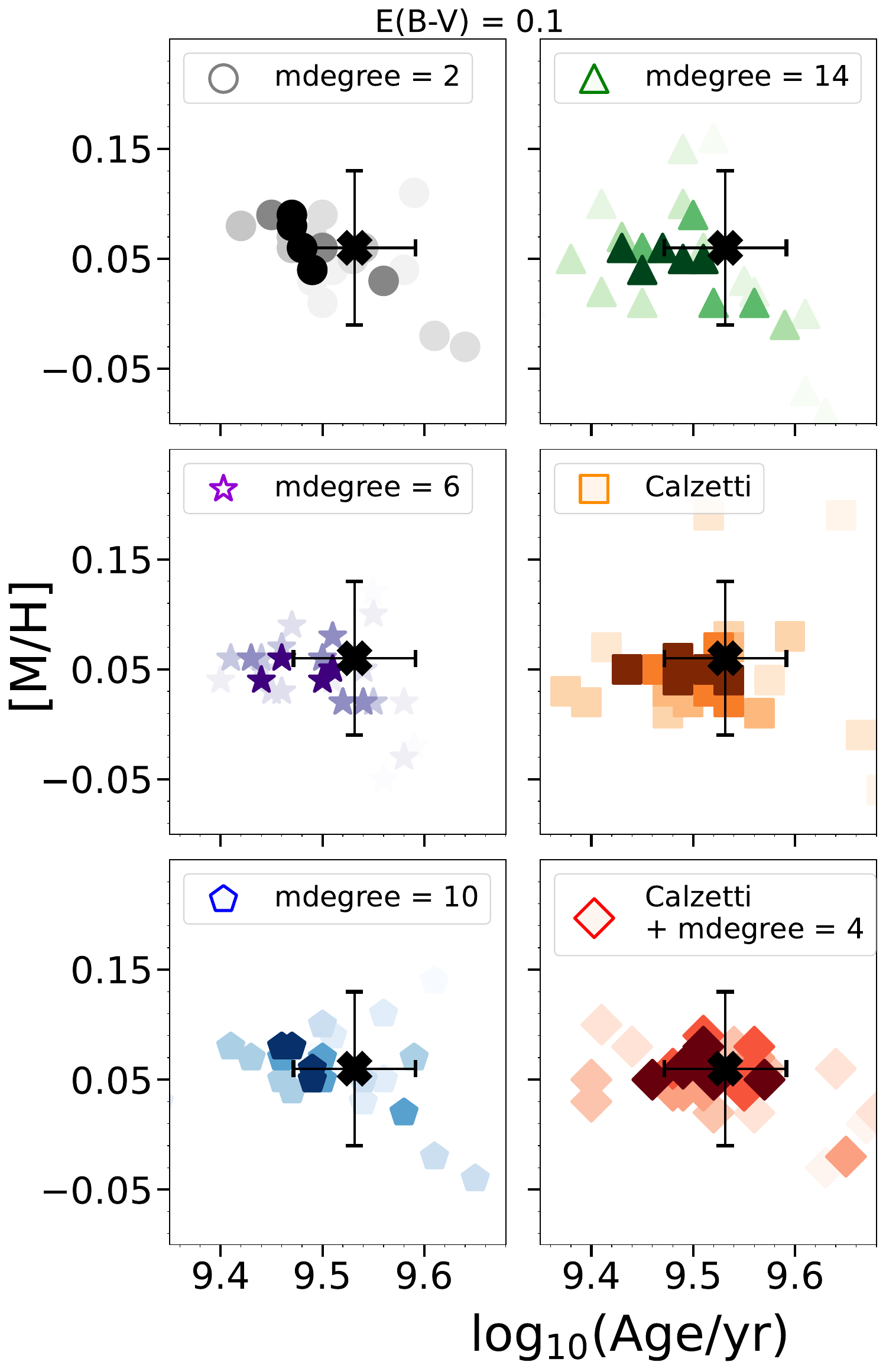}
         \caption{}
     \end{subfigure}
        \caption{Outputs from different fitting methods performed on a simulated galaxies with average age = 3.4 Gyr (= 9.53 dex) and [M/H] = 0.06 dex (black cross), not extinguished (a) and extinguished with a Calzetti reddening curve using E(B-V) = 0.1 (b). The error bars are the typical errors estimated for the age and metallicity in Appendix \ref{app:errors}, 0.07 and 0.06 dex, respectively. Fits are performed using multiplicative polynomials of degree 2, 6, 10, and 14, a Calzetti reddening curve, and a combination of a Calzetti reddening curve and multiplicative polynomials of degree = 4, as indicated in each panel. Markers represent the output values at different S/N, ranging from 5 to 100, with darker colors representing higher S/N values. }
        \label{fig:simulgal}
\end{figure}

\begin{figure*}
\includegraphics[width=\textwidth]{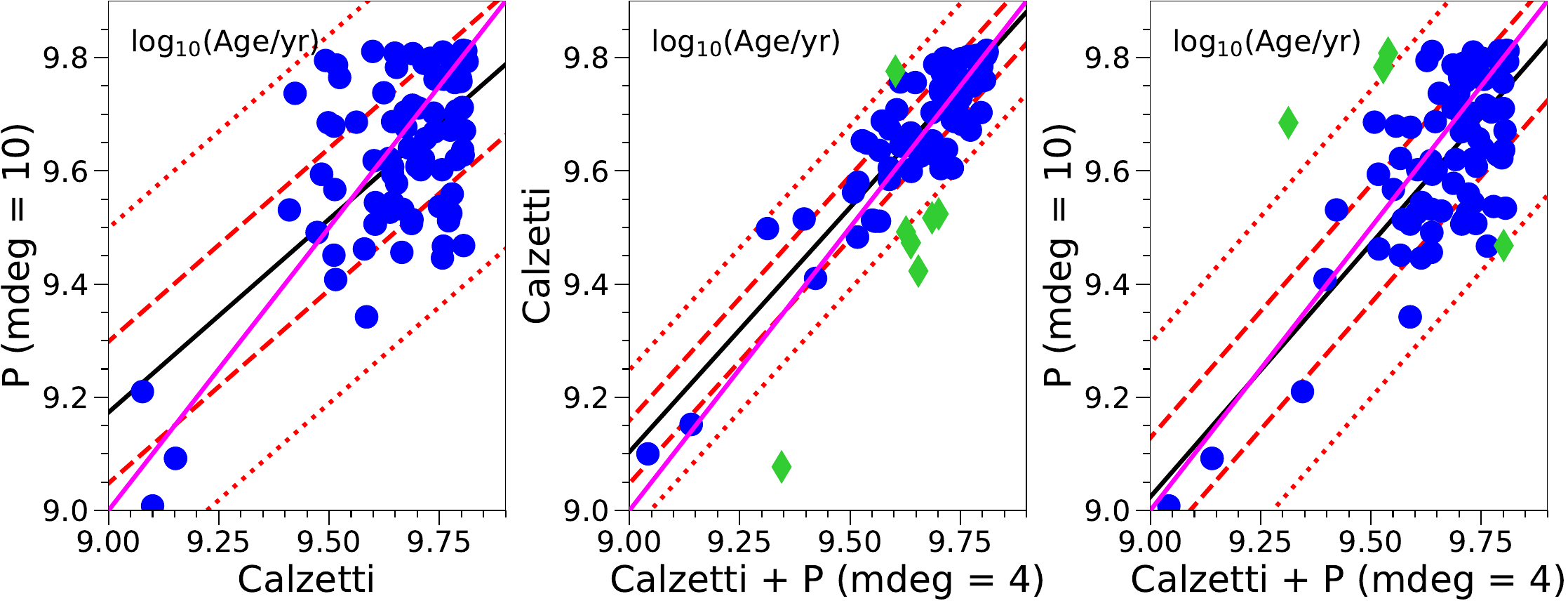}
\includegraphics[width=\textwidth]{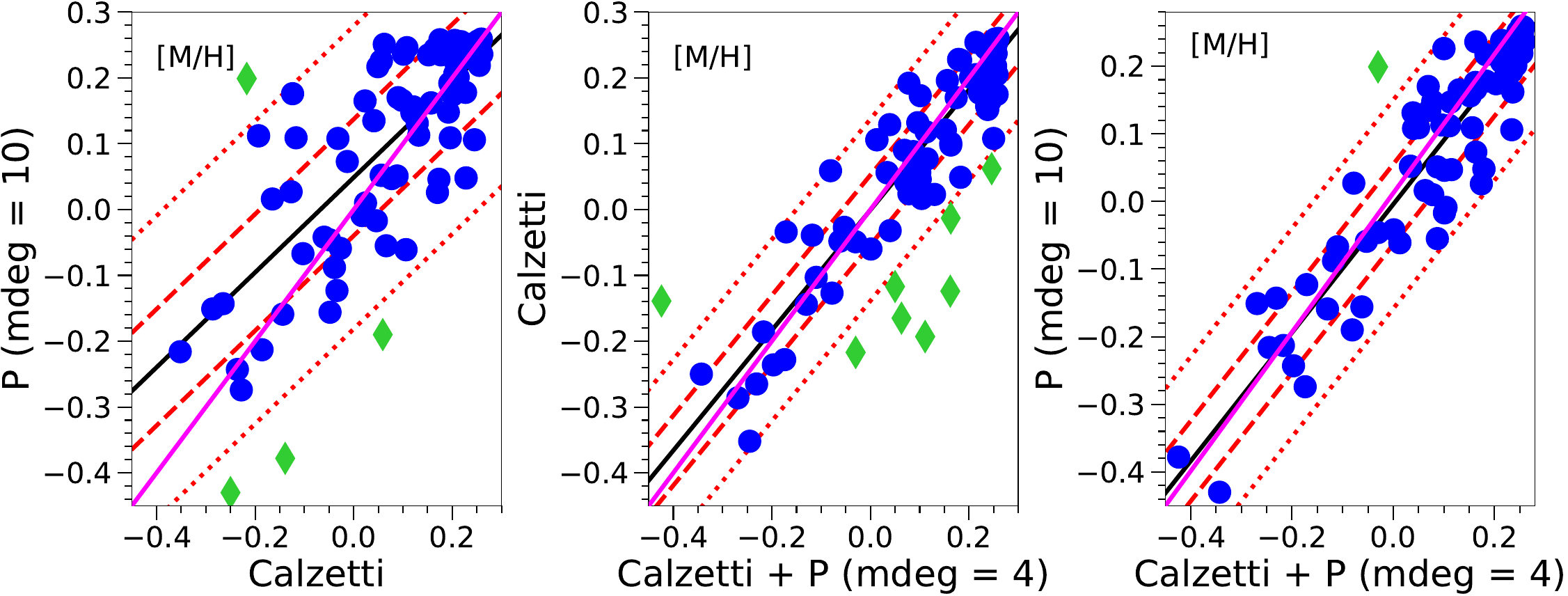}
\caption{Comparison of ages (upper panels) and metallicities (lower panels) estimated using three different fitting methods on a subsample of 64 LEGA-C galaxies with various S/N randomly chosen from our sample of quiescent galaxies from LEGA-C. The fits are performed using a Calzetti reddening law ( Calzetti ), polynomials of degree 10 ( P(mdeg = 10) ), and a combination of a Calzetti reddening law and polynomials of degree 4 ( Calzetti + P (mdeg = 4) ). In each panel, the black line is the linear fit of the estimated values, while the red dashed and dotted lines are the 1$\sigma$ and 2.6$\sigma$ lines, respectively. The blue circles are the fitted values, while the green diamonds are the outliers. The magenta line in each panel is the one-to-one relation. }
\label{fig:curvemdeg}
\end{figure*}

It is common practice, when performing full-spectral fitting, to use multiplicative polynomials to correct for inaccuracy in the flux calibration, dust extinction, and any possible problems that can affect the shape of the spectrum. Often, the order of the polynomials is chosen to be the lowest to reach convergence to a stable solution, i.e. when further increasing the degree of polynomials does not change the output values, while requiring longer computational time (e.g., \citealt{Barone+20}). However, the convergence of the solutions does not necessarily imply that the solution is optimal. Rather, it means that higher order polynomials do not `correct better' the spectrum in a statistical sense, namely the $\chi^2$ is not further reduced. 

Using high-order polynomials may not always be the optimal solution. For instance, it is known that some abundances are overestimated or underestimated by models based on MILES stars (e.g., CN, see \citealt{Vazdekis_1999, Thomas+2003}). This means that an offset between the models and the observed spectrum at the wavelengths corresponding to the absorption lines of such elements is generally expected, when using these models. However, polynomials tend to `wipe under the rug' such offsets. Since the full-spectral fitting procedure is performed over the whole spectrum, an offset of a relatively low number of spectral pixels should not affect significantly the overall fit, and using polynomials should be safe in general against small offsets. Further, for the specific case of the LEGA-C spectra, using multiplicative polynomials might be preferable. Indeed, LEGA-C spectra have been calibrated using stellar population models. This non-standard procedure could have introduced some inaccuracies that multiplicative polynomials should be able to remove.

Here, we evaluate the impact of using different orders of multiplicative polynomials to fit the stellar population properties and compare them with the fit performed using a Calzetti reddening curve. Further, since the last version of \texttt{ppxf} \citep{ppxf_2023}, it is also possible to use a combination of polynomials and a reddening curve to fit the spectra. 

To test the different methods, we simulate a spectrum of a LEGA-C galaxy combining two E-MILES SSPs with ages of 4 Gyr and 1 Gyr, respectively, and both with a metallicity of +0.06 dex. We impose that the older SSP contributes with a mass fraction of $80\%$, while the younger with $20\%$, thus resulting in an average age of 3.4 Gyr. We then match the combined spectrum to the FWHM of LEGA-C and convolve it with a gaussian with a velocity dispersion of 200 km s$^{-1}$. We then gaussianly add noise in order to achieve different values for the S/N = 5, 10, 15, 20, 25, 30, 40, 50, 75 and 100. Additionally, we perform the tests on a spectrum with the same characteristics but extinguished by a Calzetti reddening curve with E(B-V) = 0.1.

We then perform the fits as described in section \ref{sect:methods}, but using different orders of the multiplicative polynomials (2, 6, 10, 14), a Calzetti reddening curve, and a combination of a Calzetti reddening curve and multiplicative polynomials of degree 4. 

In Fig. \ref{fig:simulgal} (a) we show the results for the simulated spectra without extinction. Within the errors, all methods are generally consistent. Metallicity is well constrained in most cases, while ages are much more scattered. In particular, using polynomials generally provides more scattered outputs. Further, there seems to be some covariance between age and metallicity, and solutions are slightly biased toward young ages. Interestingly, the estimated age and metallicity do not get closer to the input values with increasing S/N, as one would expect. Using only a Calzetti curve provides similar results. Finally, solutions of the combination of a reddening curve and polynomials are slightly more scattered (still consistent within the errors), but the scatter is well distributed around the true value (seemingly, without any bias or covariance), and solutions improve when the S/N is higher, as one would expect. 

Similar results are obtained for the extinguished spectrum (b), although the polynomials are generally more scattered, especially in ages, while the fit with the reddening curve and with the combination of the curve with polynomials provide less scattered outputs, closer to the true value as the S/N increases.

We perform some of these tests on other simulated spectra with different stellar population properties. In general, we find similar results. Metallicity is generally well constrained by all methods, while ages exhibit larger scatter. Fits are generally better when the simulated spectrum is a single SSP rather than a combination of multiple SSPs, particularly for older and more metal-rich spectra. In all cases considered, a combination of a reddening curve and multiplicative polynomials better constrain the input stellar population properties.

Overall, our simulations suggest that fitting the stellar populations using a combination of a reddening curve and polynomials of order 4 provides output values closest to the input ones. For this reason, we chose this setup to fit all our spectra. We also point out that \citealt{ppxf_2023} found a similar setup to be optimal to fit the LEGA-C spectra, although the methods and templates adopted there are different from ours.

As a final check, we test different methods on real galaxies. To this aim, we consider a subsample of 64 galaxies ($10\%$ of our sample) with different redshifts and masses randomly chosen from our sample of quiescent that cover S/N from $\sim 4$ to $\sim 50$. We then perform three different fits: one is performed using only the reddening curve, another one using only multiplicative polynomials of degree 10, and the last one using both the reddening curve and multiplicative polynomials of degree 4. 

In Fig. \ref{fig:curvemdeg} we compare the age and metallicity outputs of these three fits. Here, we can see that the choice of the fitting procedure may change significantly some output, especially affecting ages. Even though it is not possible to decide a priori which is the best fitting method for such a wide range of galaxies, we note that fitting both a reddening curve and a multiplicative polynomial is somehow a compromise between the output of fitting only a reddening curve and only multiplicative polynomials.

\subsection{Dependence on the fitted spectral region}\label{app:specrange}

For a proper comparison of the stellar population properties estimated from fits of such a large variety of galaxies, it is important to perform fits to a common spectral region. On the other hand, one wants to keep the spectral range as large as possible, to better constrain the fit using a larger number of spectroscopic features and larger statistics of spectral pixels. 

For LEGA-C spectra, using a spectral region covered by all galaxies is very restrictive. This is primarily because the redshift range is relatively large, and the spectral window is relatively narrow ($\sim 1500$\AA \, wide). In our sample, the largest spectral region common to most of the galaxies including relevant spectral features to constrain the stellar population is the wavelength range $\sim 3800 - 4400 \AA$ (rest-frame). This is entirely covered by $\sim 80 \%$ of galaxies in our sample.

However, to improve the fits and better constrain the stellar population parameters, we perform the spectral fitting in the restframe wavelength region $3600 - 4600 \AA$, although less than $25\%$ of spectra cover the whole region. 

\begin{figure}
\includegraphics[width=\columnwidth]{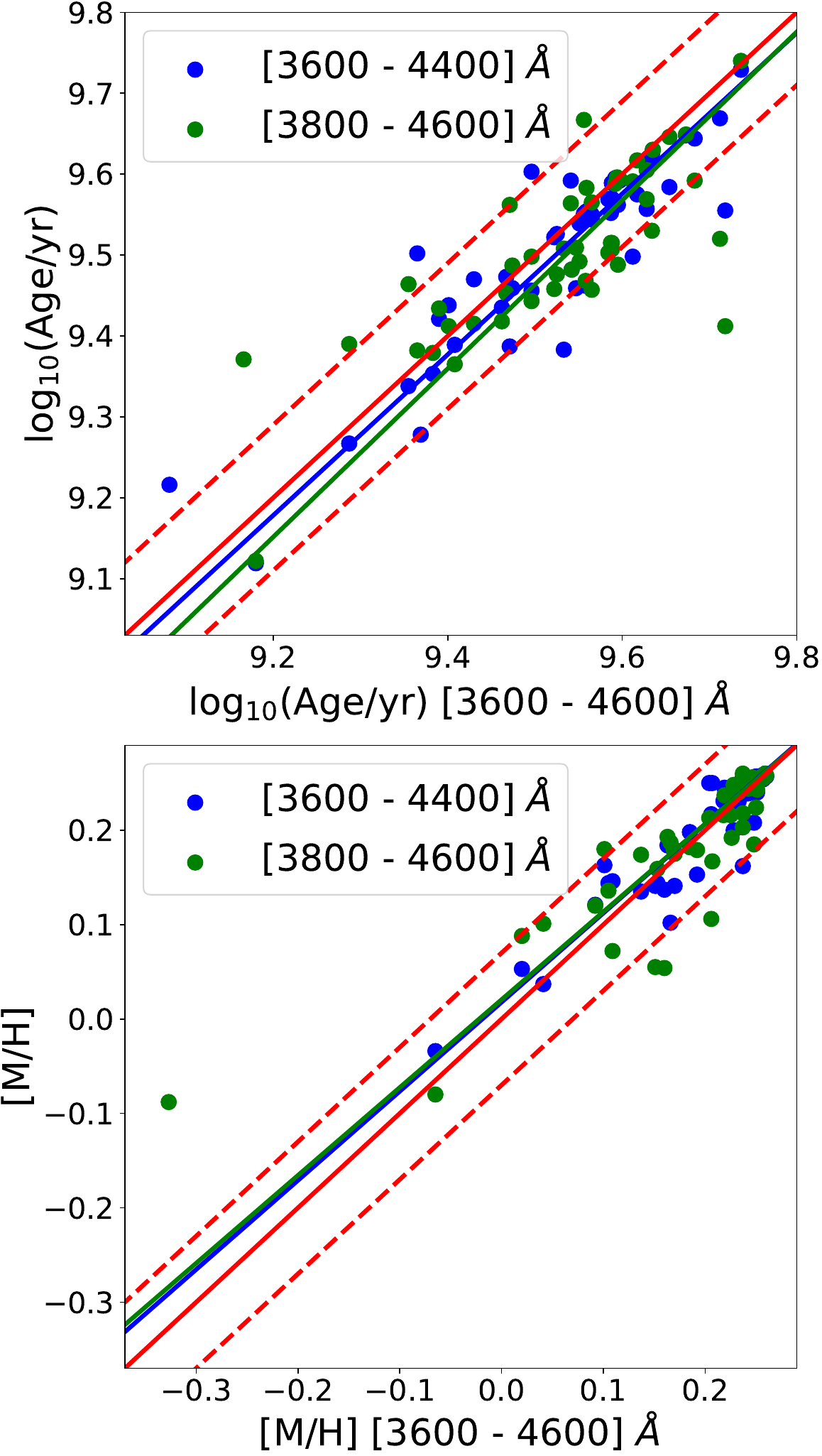}
\caption{Comparison of the ages (upper panel) and metallicities (lower panel) for fits performed over the wavelength ranges 3600 - 4600 \AA \; (x-axes) 3600 - 4400 \AA \; (blue points), and 3800 - 4600 \AA \; (red points). The blue and green lines are linear fits to the corresponding points. The red solid line is the one-to-one relation and the red dashed lines correspond to the 1$\sigma_{\scriptsize{\rm Age}}$ and 1$\sigma_{\scriptsize{\rm [M/H]}}$.}
\label{fig:fit_wrange_comp}
\end{figure}

The reason why most of the spectra do not cover the entire region is due to the different redshifts at which galaxies are observed. This means that, in most cases, galaxies that are covered up to $4600$ \AA \, (namely, those at higher redshifts) are not covered down to 3600 \AA , and vice versa. Therefore, to evaluate the impact of different spectral ranges fitted among galaxies at different redshifts, we select a random sample of 64 galaxies (10$\%$ of the total sample) covering the whole $3600 - 4600$ \AA \, spectral region, and perform the fits in this wavelength range, as well as in the restricted ranges $3600 - 4400$ \AA \, and $3800 - 4600$ \AA , and compare the outputs. 

In Fig. \ref{fig:fit_wrange_comp} we show the comparison between the ages and metallicities as fitted in the three wavelength ranges. As it is evident from the lower panel, metallicities fitted over narrower ranges show a very good agreement with those in the extended range, with an observed scatter $\approx 0.02$ dex, which is much lower than the estimated uncertainty on metallicity, $\sigma_{\mbox{\scriptsize [M/H]}} = 0.06$ dex. On the other hand, ages are generally more scattered, with a scatter of $\approx 0.06$ dex, comparable with the estimated uncertainty, $\sigma_{\log_{10}\mbox{\scriptsize Age}} = 0.07$. In all cases, the linear relations (blue and green solid lines) show very low offsets (< 0.03 dex) with respect to the one-to-one relation (red solid line), and within the $1\sigma$ scatter ($\sigma_{\mbox{\scriptsize [M/H]}} = 0.06$ dex and $\sigma_{\log_{10}\mbox{\scriptsize Age}} = 0.07$ dex, red dashed lines) we generally have a very good agreement. 

We conclude that spectra with ranges about $200$ \AA \; shorter than the fitted spectral region, $3600 - 4600$ \AA \, do not have a significant impact on the estimated stellar population properties.

Finally, 53 spectra ($\sim 8 \%$ of the sample) have minimum wavelength >4000 \AA , meaning that the characteristic `jump' of the flux at $\sim 4000$ \AA \, is not sampled in these spectra. Even though the full spectral fitting can rely on other features to estimate the stellar population parameters, not including this feature may have a significant impact on the estimated stellar population properties and, more importantly, can impair the results when compared with other galaxies. For completeness, we do not exclude these galaxies a priori, and provide their estimated ages and metallicities with a flag. We verified that including these galaxies does not affect our results at all, mainly because of their low statistics and their even distribution among different mass bins. 

\begin{figure}
\includegraphics[width=\columnwidth]{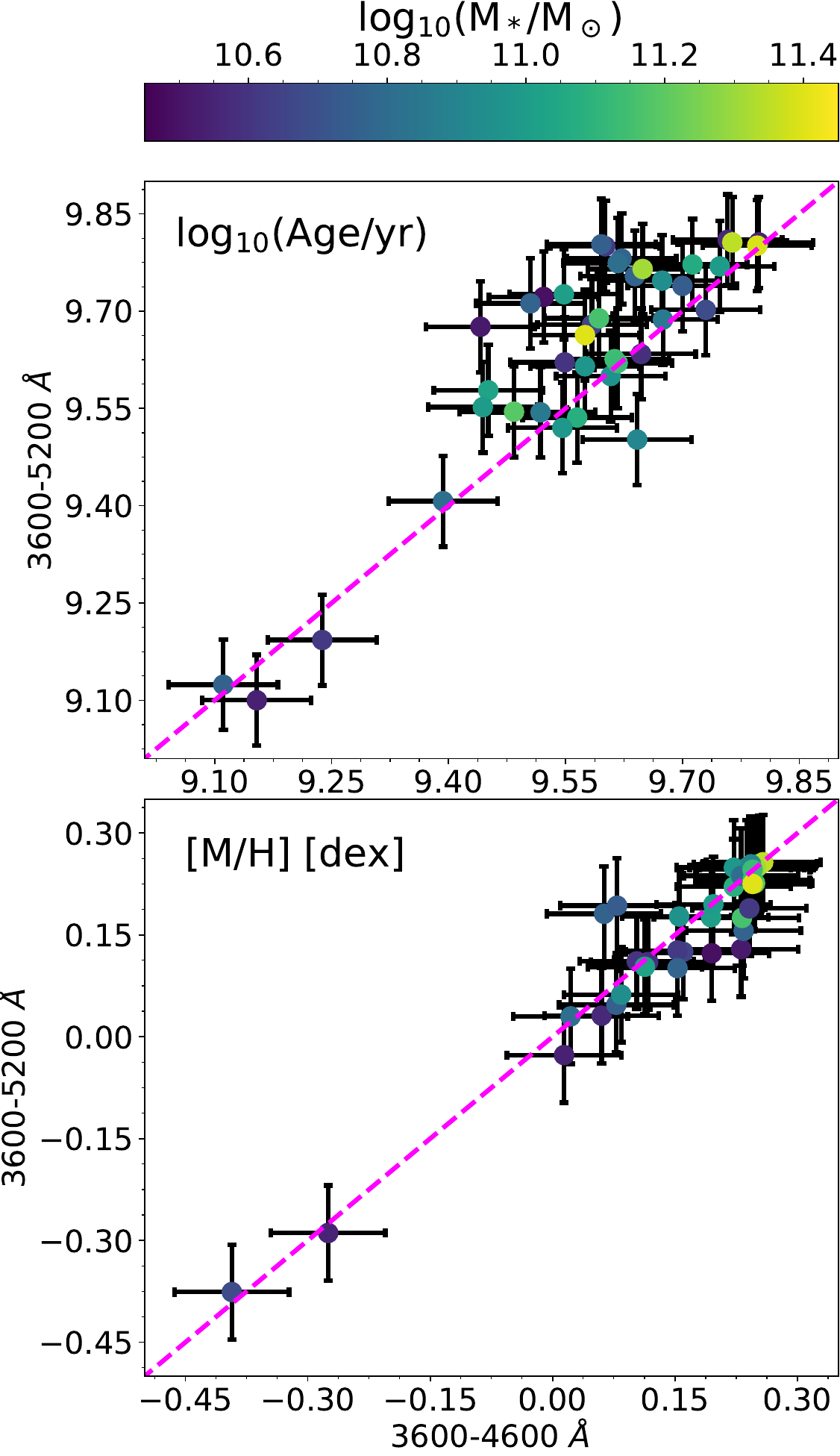}
\caption{Comparison between the estimated ages and metallicities when fitting the wavelength ranges $3600-4600$ \AA \, and $3600 - 5200$ \AA . Colors indicate the stellar masses.}
\label{fig:5200}
\end{figure}

Many galaxies in the lowest redshift bin ($0.6 \leq z < 0.7$) are observed at wavelengths up to $\sim 5200-5300$ \AA . We can thus evaluate how their stellar population parameters change when extending the fit to much redder wavelengths than the one considered so far. This is important because at these wavelengths there are many spectral features that can help better constraining the metal abundance, like magnesium and several iron lines. Further, the red part of the spectrum is more sensitive to the older stellar populations, and we may expect that the old stars would contribute more importantly to the fit.

For this purpose, we select those (40) galaxies whose spectra are measured at least up to 5200 \AA \, and down to 3700 \AA . These galaxies cover almost the entire range of stellar masses. In Fig. \ref{fig:5200} we compare their stellar population parameters estimated from fits in the two spectral regions 3600-5200 \AA \, and 3600-4600 \AA .

As mentioned above, when fitting the red hand of the spectrum, the fit is more sensitive to the older stellar populations. This can explain why many galaxies exhibit older ages. On average, fitting to 5200 \AA \, provide ages 0.06 dex older, and are thus consistent within the estimated error (0.07 dex, see Appendix \ref{app:errors}).  However, some galaxies deviate significantly from the one-to-one relation. We verified that most of these galaxies have masses log$_{10}$(M$_*$/M$_\odot) \leq 11$. These galaxies are typically composed of multiple stellar populations (see section \ref{sect:SFH}), and they are thus more sensitive to the spectral range considered.  

On the other hand, metallicities are virtually unaffected for most cases (the average metallicity is about 0.01 dex lower when fitting the extended range), and generally well consistent within the errors. There are only a few cases that deviate more than 1$\sigma$ from the one-to-one relation, and these are again all low mass (log$_{10}$(M$_*$/M$_\odot) \leq 11$ galaxies.

Overall, extending the range fitted does not have an important impact on the estimated metallicity (see also \citealt{Saracco+23} for similar results). On the other hand, ages can significantly be affected in some cases, especially at low masses, and thus affect our results. For this reason, when needed, we evaluate the impact of fitting an extended wavelength range, especially when discussing results based on age.

\section{Uncertainty estimates of age and metallicity}\label{app:errors}

To estimate the errors of age and metallicity from the full-spectral fitting, we follow the method of \cite{Barone+20}, considering the same subset of 64 galaxies used in \ref{app:mdeg_redd}. For each spectrum, we first obtain the best-fitting spectrum using the same procedure described in section \ref{sect:methods}, and thus calculate the residuals as the difference between the observed spectrum and the best fit. The residuals are then shuffled within bins about 500 \AA \, wide and added gaussianly to the best fit spectrum. Hence, we perform the fit again. On each spectrum, we repeat the shuffling and the fitting procedure 100 times. 

For each spectrum, we check the distribution of outputs from the resampled spectra. Even though these distributions are often not gaussians, they are in most cases strongly peaked at the stellar population values obtained from the best fit with narrow standard deviations, both in age and metallicity. For this reason, we take the standard deviation of ages, $\sigma_{\log_{10}\mbox{\scriptsize{Age}}}$, and metallicities, $\sigma_{\mbox{\scriptsize{[M/H]}}}$ as the errors associated to each spectrum of this subset.

\begin{figure}
\includegraphics[width=\columnwidth]{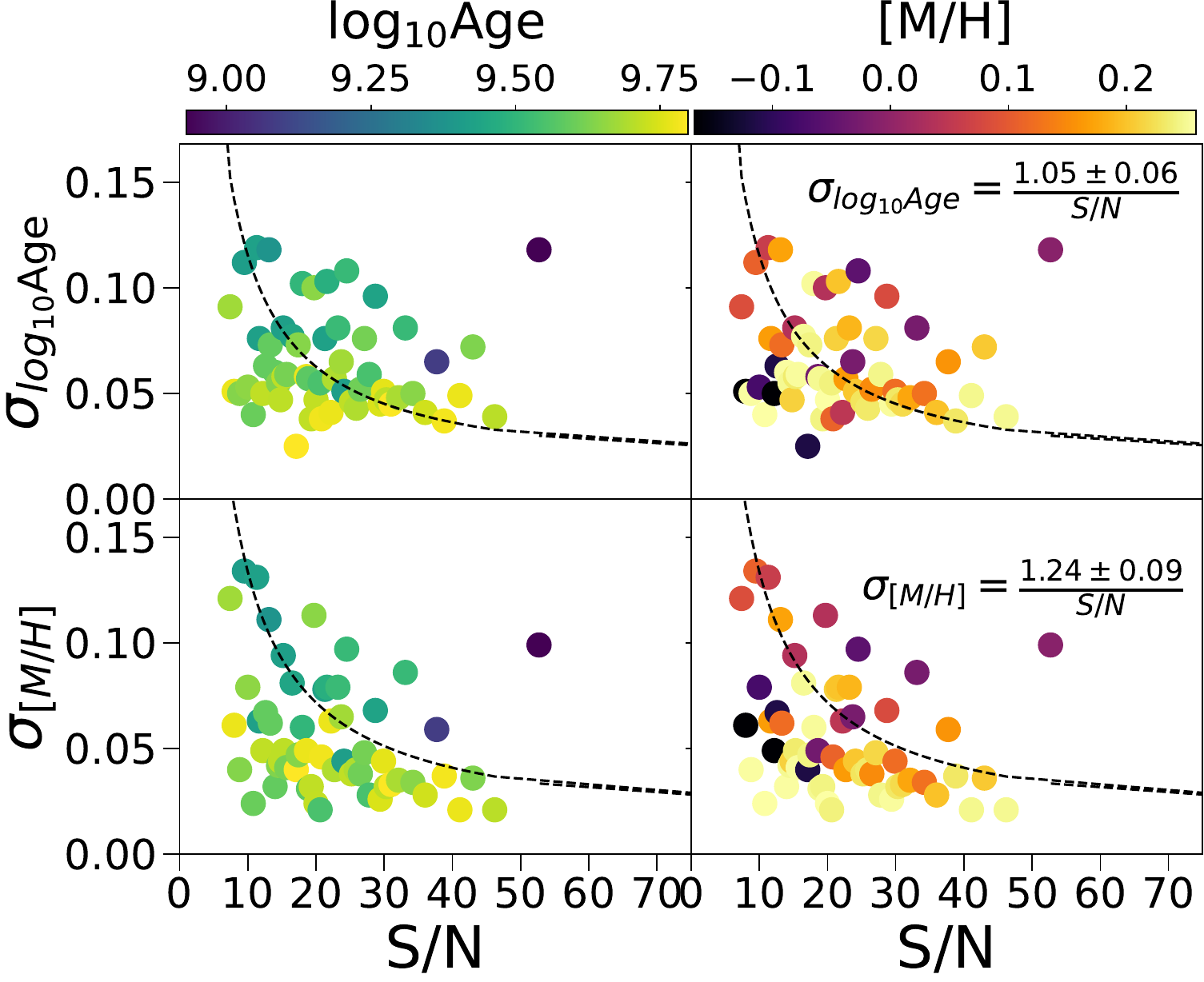}
\caption{Estimated errors on age (upper panels) and metallicity (lower panels) as a function of the S/N for a subsample of 64 galaxies. Left panels are colored in age, while right panels are colored in metallicity, as estimated from the fits. The black dashed lines represent the inverse proportionality law of errors with the S/N, whose best fit for age and metallicity is shown in the upper right and lower right panels.}
\label{fig:errors}
\end{figure}

In Fig. \ref{fig:errors} we show the distribution of the estimated errors as a function of the S/N. We then fit a relation of inverse proportionality using the Levenberg-Marquardt least-squares optimization algorithm\footnote{We used \texttt{optimize.curve\_fit} from the \texttt{Python} package \texttt{SciPy}}. Although the curve shows a slight decreasing trend of uncertainties with the S/N, the distributions are not well represented by this curve, being dominated by the scatter. Similar behavior has also been found for the mass-weighted ages and metallicities by \cite{Barone+20}. For this reason, we fix the errors of age and metallicity to the median values, namely $\sigma_{\log_{10}\mbox{\scriptsize{Age}}} = 0.07$ dex, and, $\sigma_{\mbox{\scriptsize{[M/H]}}} = 0.06$ dex, for all galaxies.

It is interesting to note, in Fig. \ref{fig:errors}, how the uncertainties on both age and metallicity decrease as the age increases. This is likely because, for older populations, a slight variation of the spectrum (due, in this case, to the resampling of the residuals) does not change the overall estimate of the stellar population properties from the fit as much as it does to younger galaxies. In this sense, older galaxies are less sensitive to the noise, and thus their properties may be in general better constrained by the fit (or, more likely, their stellar population properties are intrinsically more difficult to disentangle, also considering the limited sampling in ages and metallicities of the models).

\section{MgFe vs. [M/H]}\label{app:mgfe}

\begin{figure}
\includegraphics[width=\columnwidth]{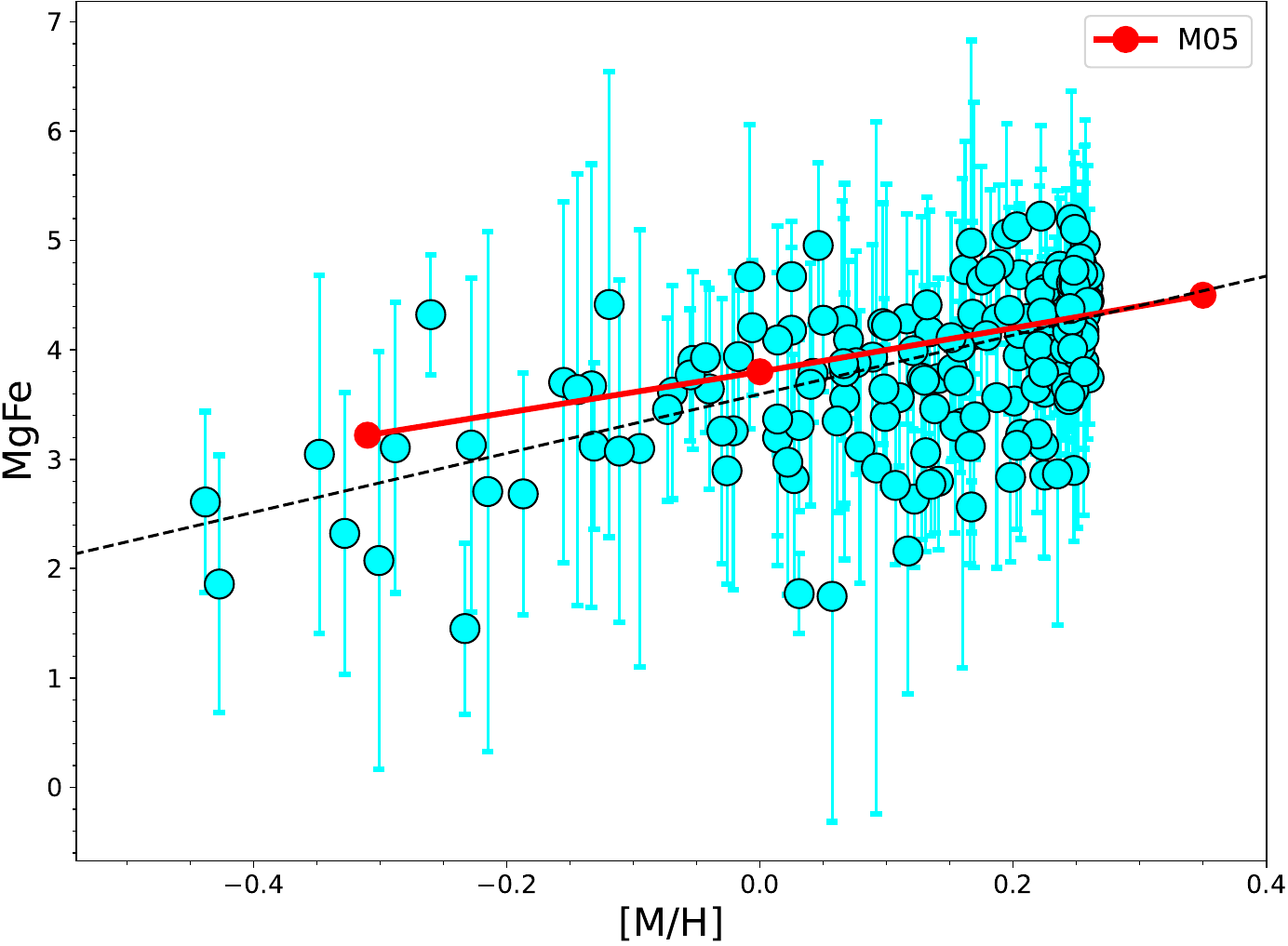}
\caption{Correlation between MgFe and [M/H]. The turquoise circles are observed MgFe values and corresponding errors at the metallicities estimated from the fits (section \ref{sect:methods}). The black dashed line is the linear fit of the data. The red line shows the predictions of M05 models for the MgFe index at different metallicities (red circles).}
\label{fig:mgfe_met}
\end{figure}

In Fig. \ref{fig:mgfe_met} we show the correlation between the metallicity, [M/H], estimated from the fits, and the MgFe index, introduced in section \ref{sect:meme}. Notwithstanding the uncertainties on MgFe, and the limit to [M/H] of the models, there is a clear linear relation between the index and the metallicity, as highlighted by the linear fit (black dashed line).

To address the problem of the limits of the E-MILES models and to further assess the relation between MgFe and [M/H], we consider the \citet{Maraston2005} (M05) models, reaching higher metallicities than E-MILES. Specifically, we considered M05 models with an age of 4 Gyr, corresponding to the average age of our LEGA-C galaxies, and with metallicities up to 0.35 dex. We show the predictions of M05 for MgFe in Fig. \ref{fig:mgfe_met} with a red solid line. For M05 models, the relation between [M/H] and MgFe is clearly linear and consistent with the estimates from the LEGA-C spectra.

We conclude that MgFe is a good tracer for metallicity.

\section{Tests on MEH }\label{app:MEH}

\begin{figure}
\includegraphics[width=\columnwidth]{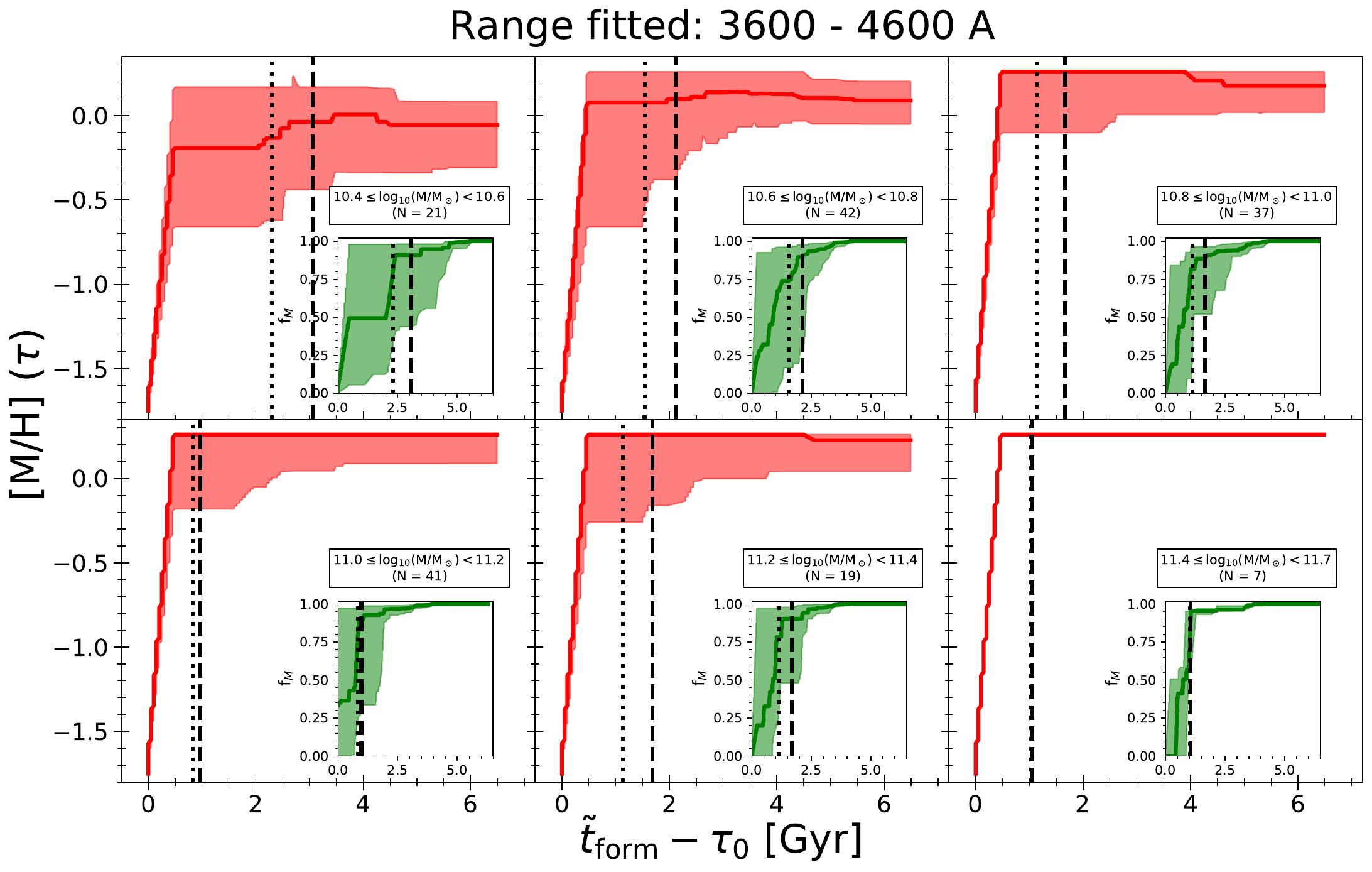}
\includegraphics[width=\columnwidth]{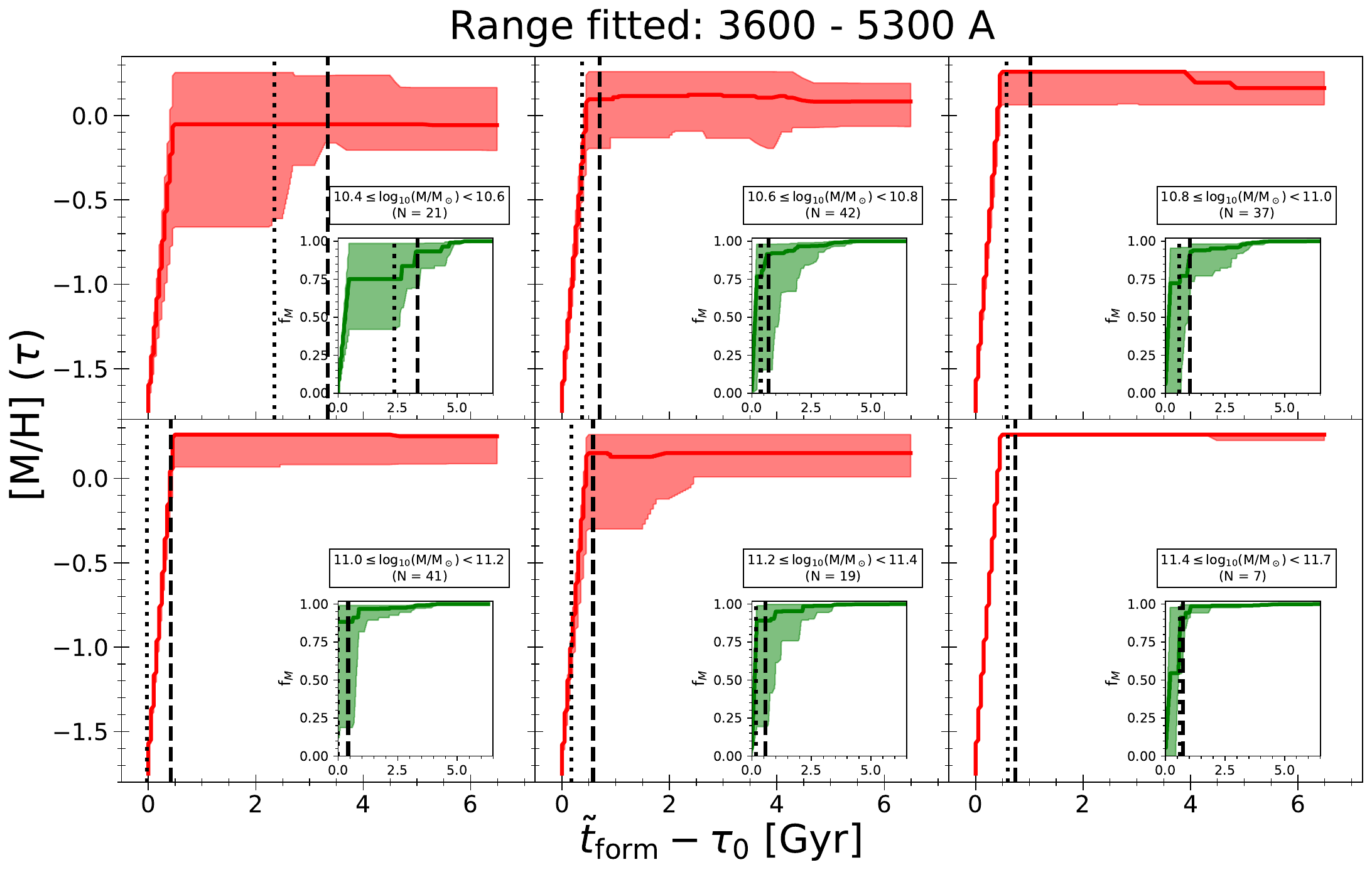}
\includegraphics[width=\columnwidth]{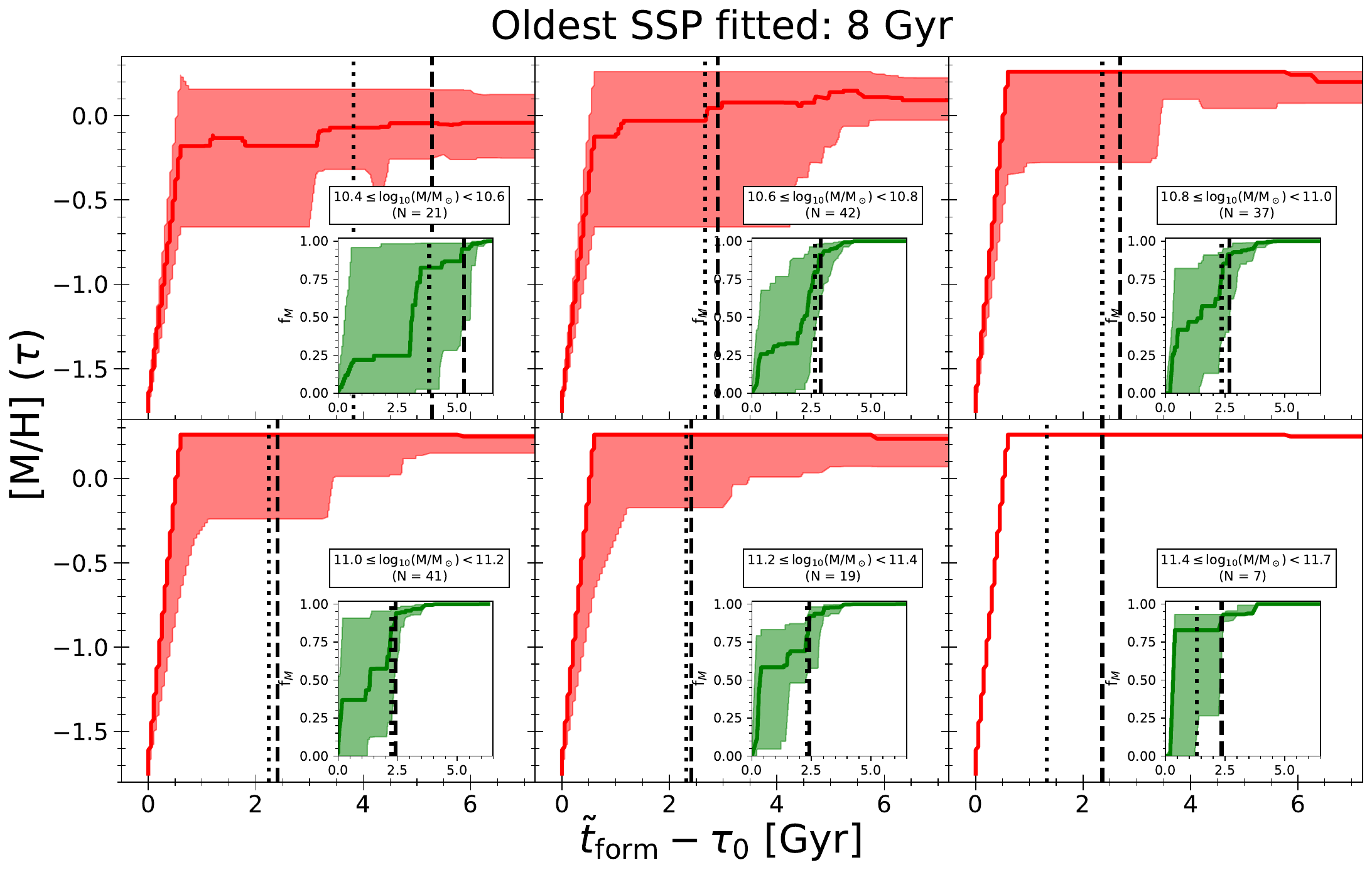}
\caption{MEH and SFH curves at different mass bins for a subsample of 183 galaxies whose spectra are observed at $\lambda > 5200 \AA$. In the upper figure, fits are performed as described in section \ref{sect:sfh_methods}. In the middle figure, fits are performed to the extended range 3600-5300 \AA . In the lower figure, fits are performed including input SSPs up to 8 Gyr old.}
\label{fig:app_MEHrange}
\end{figure}

Here we discuss some tests performed to understand how MEH and SFH curves change when fitting an extended wavelength range, when including older SSPs, and how they are affected by S/N. 

As discussed in Appendix \ref{app:specrange}, fitting a different spectral range can affect the output of \texttt{pPXF}, especially when extended to longer wavelengths. To evaluate the impact on MEHs and SFHs we consider a subsample of 167 galaxies ($\sim 26\%$ of the total sample), for which the DR3 provides measurements of the spectral index Mgb, meaning that we have a spectral coverage extended until up $\lambda \approx 5200$ \AA . We perform the fits on the extended wavelength range (up to $\lambda = 5300$ \AA), and construct the MEHs and SFHs. In Fig. \ref{fig:app_MEHrange} we show the difference of the curves for the same subsample, divided into mass bins, as fitted in the range $3600-4600$ \AA \, and $3600-5300$ \AA .

In general, there is no substantial difference when fitting different wavelength ranges. The only noticeable differences are in the two lowest mass bins. In particular, in the lowest mass bin, the median curve is flatter, reflecting the steeper median SFH, while the envelopes are pretty similar. Vice versa, in the other mass bin the median trend is essentially the same, but the percentiles' envelope is slightly narrower (within the first 2 Gyr). These secondary differences likely reflect the fact that when extending the wavelength range we are putting more weights on the older population. Also, note that the statistics are significantly lower. In any case, the qualitative results are in general the same.

We then investigate how including older SSPs to fit the spectra affects the results. Indeed, even though we verified that the general distribution of the average age and metallicities (calculated using equations \eqref{eq:age} and \eqref{eq:met}) are the same when fitting with older SSPs, the curves, which, differently from the weighted averages, are sensitive to the weights assigned by the fit to each SSP, could in principle vary significantly. For this reason, we consider the same subsample as above and fit it including SSPs up to 8 Gyr old\footnote{Note that this is older than the age of the Universe at the lowest redshift considered here, $z=0.6$, i.e. $\sim 7.8$ Gyr.}. The lowest panels of Fig. \ref{fig:app_MEHrange} show the results of this fit. The SFHs are generally larger and the $\tau_{90}$ values are typically older than those found with younger templates, and it reflects generally larger SFHs. However, the difference decreases as the mass increases, and the general qualitative trends are the same. Interestingly, MEH curves are not affected as much as SFHs, and the only difference is that the envelopes are slightly larger.

\begin{figure}
\includegraphics[width=\columnwidth]{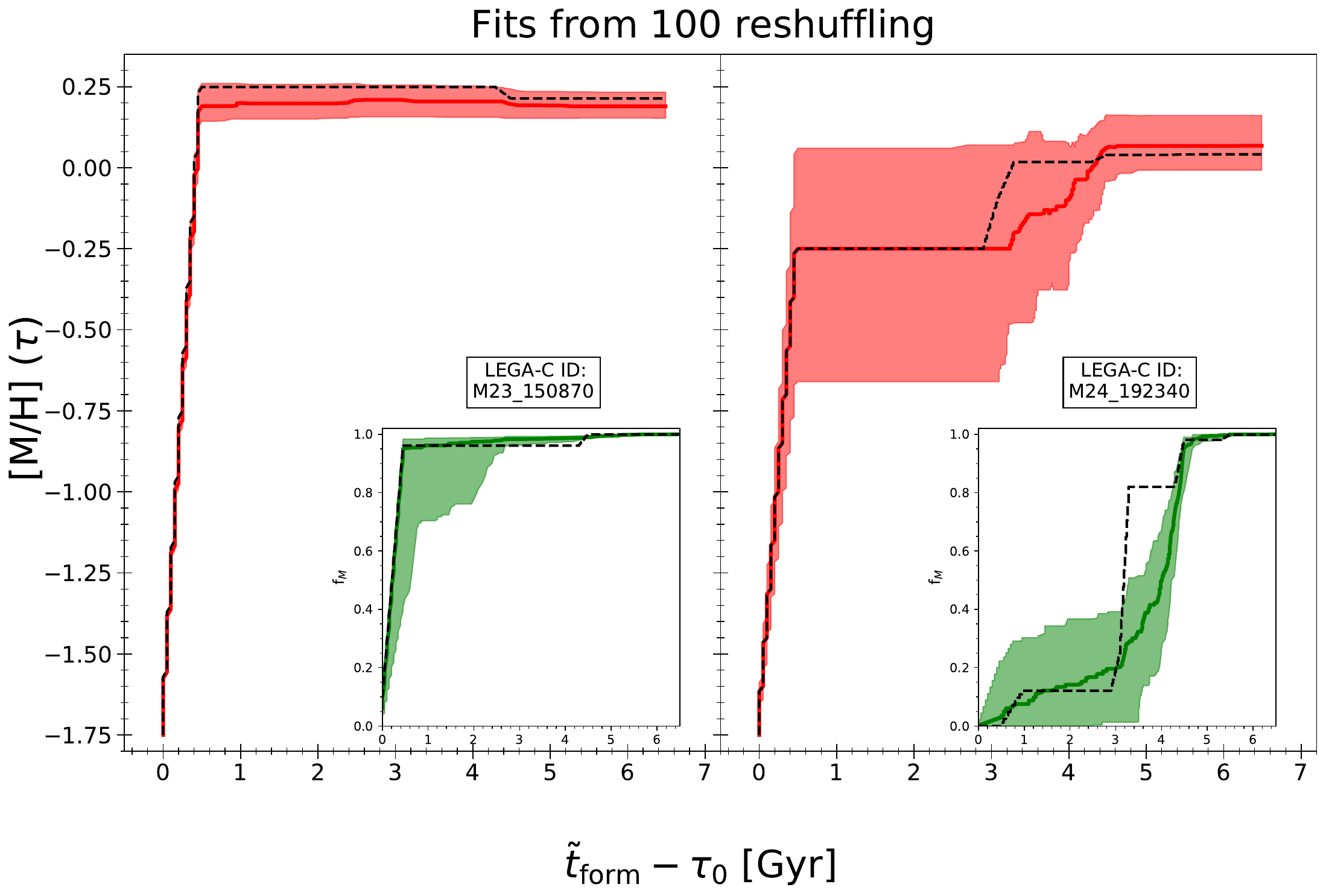}
\caption{MEH and SFH median curves and 16th-84th percentiles resulting from the fits of two galaxies with $\Delta\tau \leq 1$ Gyr (left) and $\Delta\tau > 1$ Gyr (right), whose best-fit spectra have been fitted 100 times after the reshuffle of the residuals and gaussianly adding them to it.}
\label{fig:app_MEHshuffle}
\end{figure}

As a final test, we evaluate how curves are affected by the noise. To this end, we select six galaxies, three with $\Delta\tau \leq 1$ Gyr, and three with $\Delta\tau > 1$ with increasing metallicity, with S/N between 10 and 30 (the average S/N in our sample is about 21). For each galaxy, we reshuffle the noise 100 times, in the same way we did to estimate the errors on the age and metallicity (see Appendix \ref{app:errors}), add it gaussianly to the best-fit spectrum, and perform again the fit. In Fig. \ref{fig:app_MEHshuffle} we show two examples of two galaxies with $\Delta\tau \leq 1$ Gyr and $\Delta\tau > 1$, both with S/N$\sim 20$. The envelopes represent the 16th-84th percentiles of the 100 realizations. As it is evident, the MEH and SFH of the galaxy with $\Delta\tau \leq 1$ (left panel) are essentially always the same. The galaxy with $\Delta\tau > 1$ shows a larger scatter, but the general trend is unchanged, and the galaxy always increases its metallicity in time. This indicates that part of the scatter we observe in Fig. \ref{fig:MEH} is rising from uncertainty on the fit, but the qualitative trends are trustworthy. Similar results are found for galaxies with lower and higher S/N.

\section{Predictions from simulations}\label{app:simulations}
\begin{figure}
\includegraphics[width=\columnwidth]{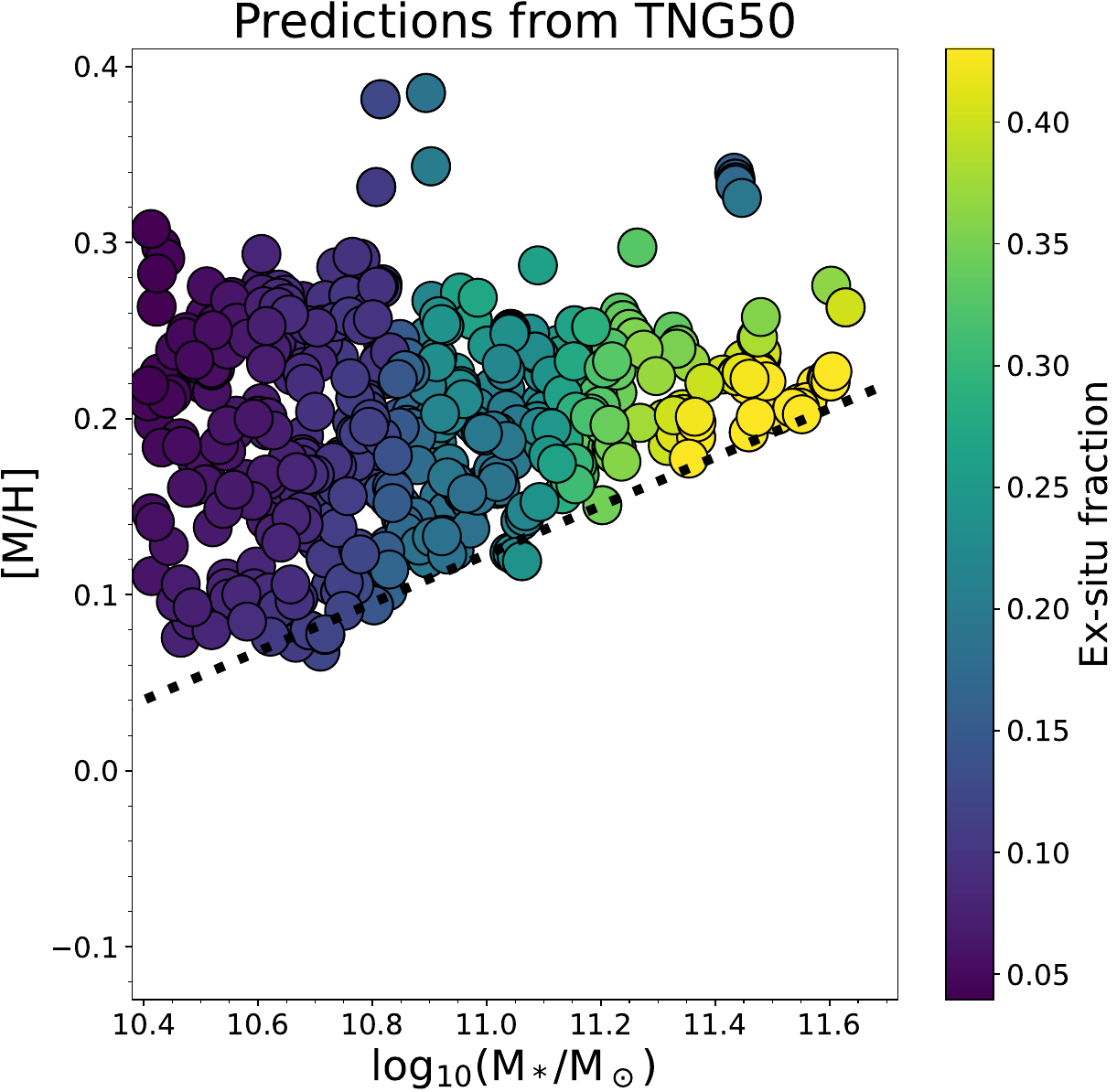}
\caption{Comparison between observed LEGA-C (grey) and simulated TNG50 (colored) quiescent galaxies. The color code of TNG50 is the fraction of ex-situ (i.e. accreted) stars. Colors are smoothed with \texttt{LOESS} (\texttt{frac=0.2}). The magenta line is the MEME relation. The black dotted line is the MEME relation calculated for TNG50 galaxies using equation \eqref{eq:meme}.}
\label{fig:tng_legac}
\end{figure}
In this Appendix, we investigate the metallicity-mass diagram as predicted by the TNG50  hydrodynamical simulations \citep{Pillepich+19, Nelson+19}, and by the GAlaxy Evolution and Assembly (GAEA) semi-analytical model \citep{DeLucia+14, Hirschmann+16, DeLucia+24}. We caution the reader that comparing observations and simulations is not a straightforward task, and a proper comparison requires appropriate tests. However, this is beyond the scope of this paper. The following comparisons are only intended to be qualitative.

We select quiescent galaxies from TNG50, mimicking the selection of LEGA-C galaxies presented in this work. In detail, the initial selection of TNG50 galaxies is similar to the ones observed by the LEGA-C survey \citep{Wu+21}. We then additionally constrain the selection to quiescent galaxies only using flags from \citet{Pillepich+19} (see their Table 1) and apply the same mass completeness cuts with redshift, as described in Section \ref{sect:data}. Metallicities are calculated within twice the half-mass radius. In Fig. \ref{fig:tng_legac}, we show the distribution on the metallicity-mass diagram of the TNG50 galaxies, color-coded by the fraction of ex-situ (accreted) stars (these will be presented in Boecker et al., in prep.). 

In agreement with observations, TNG50 simulations show that while lower-mass galaxies span a larger range of metallicities, higher-mass galaxies are all metal-rich. We note that the absolute values of the metallicities in TNG50 are typically higher than the ones estimated from fits, and there are no metallicities as low as the ones estimated from the observed spectra, which is likely a combined effect of higher input metallicities in TNG50 and an inflated scatter due to S/N in observations (see Boecker et al., in prep.). Notwithstanding these differences, simulations clearly show a mass-dependent lower limit resembling the MEME relation is spottable, although the slope is flatter. Notably, similar to observations, lower-mass galaxies have a larger scatter, and they reach metallicities as high as the most massive galaxies.

\begin{figure}
\includegraphics[width=\columnwidth]{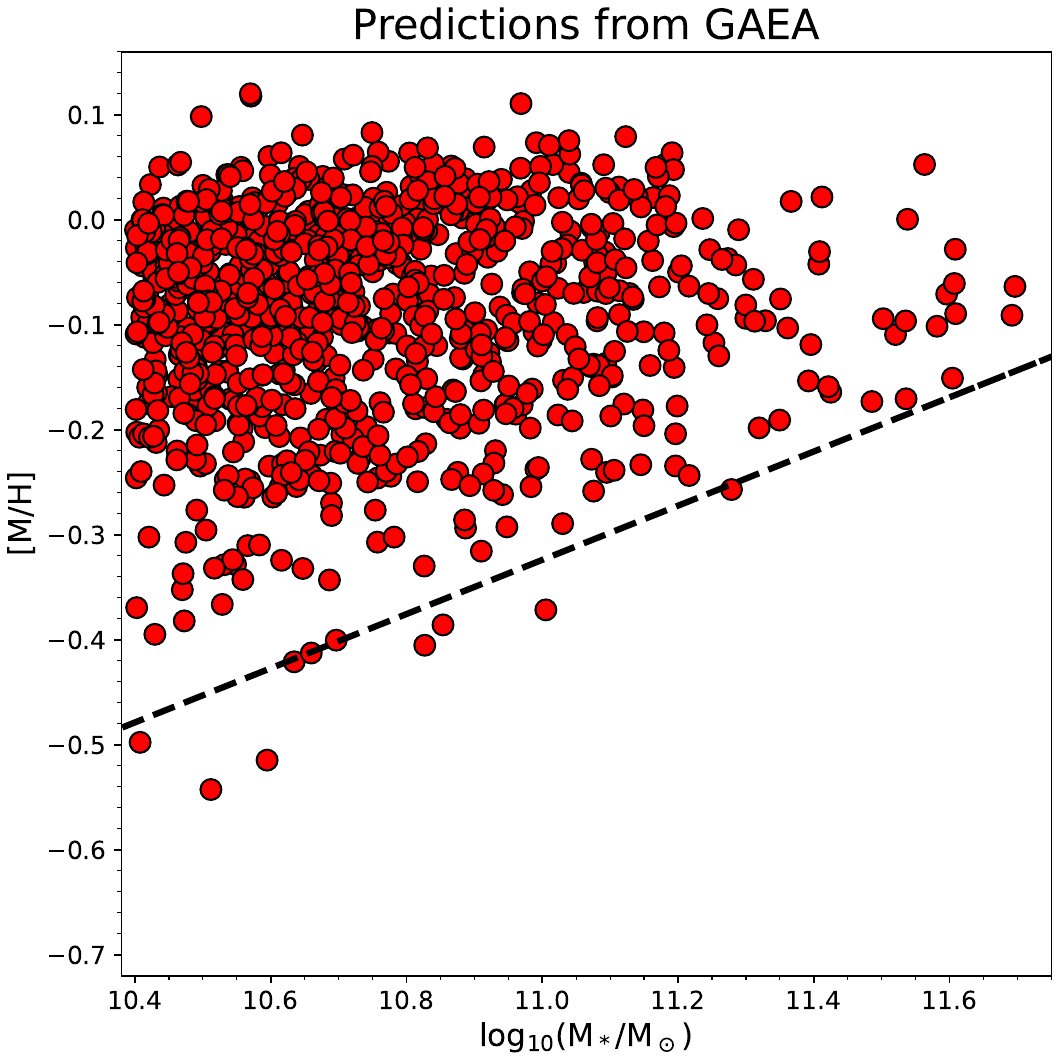}
\caption{Metallicity-mass diagram as predicted by the GAEA semi-analytical models. The black dashed line is the MEME relation calculated for the GAEA predictions.}
\label{fig:gaea}
\end{figure}

As the colors clearly indicate, the fraction of ex-situ stars increases with the stellar mass. 
Interestingly, lower-mass galaxies have the lowest fractions of ex-situ stars, and thus most of stars are formed in situ. This implies that the larger scatter in metallicity is not due to mergers affecting the distribution.

To have an independent comparison with simulations, we consider the GAEA semi-analytical model. GAEA is a state-of-the-art theoretical model that simulates galaxy formation and evolution in cosmological volumes. Here, we consider the predictions from the latest renditions of the GAEA model, presented in \citet{DeLucia+24}\footnote{For further information about the model, see also \url{https://sites.google.com/inaf.it/gaea}}.

As a representative sample of the LEGA-C galaxies, we consider a sub-volume (about 10$\%$) representative of the Millennium Simulations \citep{millennium}, and select 650 quiescent galaxies from the predicted UVJ diagram at $z\sim 0.8$, with the same mass distribution of our sample. We note that, due to the different methods for estimating the metallicities, a direct comparison with LEGA-C measurements would be incorrect. In particular, the GAEA model predicts systematically lower metallicities than the observed ones (e.g., \citealt{Saracco+23}).

In Fig. \ref{fig:gaea} we show the metallicity-mass diagram as predicted by the GAEA model. Similar to observations, GAEA predicts a mass-dependent lower limit to metallicity. The predicted lower limit is flatter than the observed one, analogously to the TNG50 simulations. Finally, lower-mass galaxies span a larger range of metallicities than most massive galaxies, reaching comparable, high metallicities.

We conclude that both the TNG50 hydrodynamical simulations and the GAEA semi-analytical model simulations predict the mass-dependent lower limit to metallicity observed for real galaxies. They both also predict that lower-mass galaxies span larger metallicity ranges, and reach values as high as the most massive galaxies. Note that, differently from observations, the scatter in the simulations is not affected by S/N. Being TNG50 and GAEA completely independent, the observed behavior is likely related to the common physics driving the metallicity evolution in simulations, regardless of the different approaches and recipes adopted.


\end{appendix}

\end{document}